\documentclass[twocolumn,english,aps,prb]{revtex4}
\usepackage[T1]{fontenc}
\usepackage[latin9]{inputenc}
\usepackage{graphicx}

\providecommand{\tabularnewline}{\\}

\usepackage{hyperref}
\def\urlprefix{}
\def\url#1{} 

\usepackage{babel}

\begin{document}

\title{Mottness on a triangular lattice}

\author{Dimitrios Galanakis$^{1}$, Tudor Stanescu$^{2}$, Philip Phillips$^{1}$}

\affiliation{$^{1}$Loomis Laboratory of Physics, University of Illinois at Urbana-Champaign,
1100W. Green St., Urbana, IL., 61801-3080}

\affiliation{$^{2}$Condensed Matter Theory Center, Department of Physics, University
of Maryland, College Park, Maryland 20742-4111}
\begin{abstract}
We study the physics on the paramagnetic side of the phase diagram
of the cobaltates, $Na_{x}CoO_{2}$, with an implementation of cellular
dynamical mean field theory (CDMFT) with the non-crossing approximation
(NCA) for the one-band Hubbard model on a triangular lattice. At low
doping we find that the low energy physics is dominated by a quasi-dispersionless
band. At half-filling, we find a metal-insulator transition at $U_{c}=5.6\pm0.15t$
which depends weakly on the cluster size. The onset of the metallic
state occurs through the growth of a coherence peak at the chemical
potential. Away from half filling, in the electron-doped regime, the
system is metallic with a large, continuous Fermi surface as seen
experimentally. Upon hole doping, a quasi non-dispersing band emerges
at the top of the lower Hubbard band and controls the low-energy physics.
This band is a clear signature of non-Fermi liquid behavior and cannot
be captured by any weakly coupled approach. This quasi non-dispersive
band, which persists in a certain range of dopings, has been observed
experimentally. We also investigate the pseudogap phenomenon in the
context of a triangular lattice and we propose a new framework for
discussing the pseudogap phenomena in general. This framework involves
a momentum-dependent characterization of the low-energy physics and
links the appearance of the pseudogap to a reconstruction of the Fermi
surface without invoking any long range order or symmetry breaking.
Within this framework we predict the existence of a pseudogap for
the two dimensional Hubbard model on a triangular lattice in the weakly
hole-doped regime.

\end{abstract}
\maketitle

\section{Introduction}

Charge carriers in the cobaltates, $Na_{x}CoO_{2}$, are located in
two dimensional $CoO_{2}$ layers separated by insulating layers of
{\small $Na^{+}$ ions which act as electron donors. Their structure
is a triangular net of edge-sharing oxygen octahedra with the $Co$
atoms occupying the center and the $Na$ atoms playing the role of
electron donors. The octahedral symmetry around the $Co$ ions results
in a splitting of the $d$-orbitals in two $e_{g}$ and three lower
lying $t_{2g}$ orbitals. The trigonal distortion of the }$CoO_{2}$
layers further splits the $t_{2g}$ orbitals into one $a_{1g}$ and
two lower $e_{g}^{\prime}$ \citet{Singh_LDA_Cobaltate}. The valence
of the cobaltate ions is $Co^{4-x}$ which means that the Fermi surface
will lie in the $a_{1g}$ orbital which will range from half to fully
filled. Consequently, the cobaltates constitute a realization of strongly
correlated electron physics on a triangular lattice. Across their
phase diagram they exhibit a wide range of behavior {\small \citep{foo_NCOPhase}
}ranging from a paramagnetic Fermi liquid at low $Na$ concentration
$x$, a strange Fermi liquid with Curie Weiss magnetic susceptibility
for high $x$ and a singular insulating state at $x=0.5$. While the
paramagnetic metal exhibits some properties akin to that of a Fermi
liquid, the cobaltates still remain strongly correlated systems. For
example, experiment{\small{} \citep{HasanARPESCobaltates}} and theory
{\small \citep{lee_picket_Cobaltates}} place the hopping matrix element
and the on-site repulsion at $t=0.2eV$ and $U=4eV$, respectively.

Various ARPES studies have been performed \citep{qian_QP_Instabilities_ARPES,yang_ARPES_cobaltates,qian-quasiparticle},
which suggest a Fermi surface which consists only of a large $a_{1g}$
hole pocket with the $e_{g}^{\prime}$ orbitals lying under the Fermi
surface. This is in contrast with LDA calculations \citet{Singh_LDA_Cobaltate}
which suggest the existence of peripheral $e_{g}^{\prime}$ hole pockets.
In an effort to resolve this discrepancy, several multi-band DMFT
studies have been performed \citep{ishida_cobaltates_DMFT,marianetti_cobaltates}.
Finally a few CDMFT calculations \citet{kyung_CDMFT_Triangle,lee_monien_dual_fermion}
address mostly the Mott transition on a triangular lattice or compare
different impurity solvers.

Motivated by the cobaltates, we address in this paper, the properties
of strong electron correlation on a triangular lattice. Of particular
interest will be the nature of the Mott transition at half-filling
on such a lattice. A triangular lattice offers an ideal playground
for exploring the Mott transition as a result of the inherent magnetic
frustration that is present. We find that a critical value of $U=5.7t$
separates paramagnetic insulating and metallic phase. Away from half-filling
we find a metallic phase with a large Fermi surface as is observed
experimentally.

This paper is organized in three main sections. In section \ref{sec:Description}
we give an overview of the computational scheme, the cluster dynamical
mean field theory with the non-crossing approximation as the impurity
solver in the context of the one-band Hubbard model. In section \ref{sec:Limitations}
we discuss the the two main issues related to the consistency of the
method: the proper periodization procedure to obtain physically correct
lattice quantities and the cluster size dependence of the the results.
We show that, for a given small cluster size, the method breaks down
at certain filling values and we argue that the cluster size independence
should be the ultimate consistency criterion. In section \ref{sec:Results}
we present the results of the simulation. In \ref{sub:Dispersionless-low-energy},
we show the existence of a quasi-dispersionless low energy band, which
is signature of strong correlations and incompatible with Fermi-liquid
physics. In \ref{sub:Mott-transition}, we discuss the Mott transition
on the triangular lattice. Finally, in \ref{sub:Pseudogap}, we argue
for the existence of a pseudogap at low hole doping.

\section{Description of the method\label{sec:Description}}

We start with the one-band Hubbard model,\begin{equation}
H=-t\sum_{\sigma,\left\langle i,j\right\rangle }c_{i\sigma}^{\dagger}c_{j\sigma}+c.c.+U\sum_{i}n_{i\downarrow}n_{i\uparrow}\label{eq:HubbardModel}\end{equation}
where $t$ is the matrix element for hoppings between nearest neighbor
sites, $\langle i,j\rangle$, and $U$ is the on-site repulsive interaction.
We assume that the single-band Hubbard model on a triangular lattice
captures the main features of strongly correlated physics in the presence
of magnetic frustration. We restrict our study to the paramagnetic
state\begin{equation}
\left\langle n_{i\downarrow}\right\rangle =\left\langle n_{i\uparrow}\right\rangle =n/2\label{eq:Fillings_Equal}\end{equation}
which is consistent with the experimental observations in cobaltates
for $x<0.5$.

As a computational tool, we use in our investigation a real-space
cluster generalization of dynamical mean field theory (DMFT). \citep{GeorgesReview}
The DMFT has been a very successful tool in investigating many aspects
of strongly correlated systems. In this method one single site is
treated as an impurity embedded in an effective bath consisting of
the rest of the sites the properties of which are captured by the
hybridization function. It is exact in infinite dimensions or more
precisely in infinite coordination number $z$ and it can successfully
describe the antiferromagnetic order. However for many applications
it is necessary for the short range (few lattice site) correlation
to be described accurately. In the cluster DMFT method (CDMFT), \citep{Kotliar_CDMFT}
a cluster extending in a small number of sites is treated as the impurity
and therefore the local (cluster) degrees of freedom are treated exactly.
The rest of the lattice, the bath, is described by a multi-component
hybridization function. 

All cluster-DMFT-based algorithms contain the following major components. 
\begin{itemize}
\item An impurity solver, which evaluates the cluster Greens function from
the hybridization function.
\item A self consistency condition which expresses the hybridization function
with respect to the cluster Greens function.
\item A periodization procedure which connects the lattice quantities with
the cluster quantities.
\end{itemize}

\subsection{Impurity solver\label{sub:Impurity-solver}}

The impurity solver evaluates the cluster Green's function, given
the {}``external'' hybridization function. Various impurity solvers
have been proposed in the literature such as exact diagonalization
and quantum Monte-Carlo. However those can only be implemented in
imaginary time and an analytic continuation is required to obtain
real time properties. A real time impurity solver is the non-crossing
approximation (NCA), which is a first order perturbation theory with
respect to the hybridization function. It has the advantage of being
very fast and relatively easy to implement. The NCA has been a valuable
tool for extracting the physics of the Anderson impurity models. The
NCA equations can be obtained by using the slave boson method \citep{NCAColeman}
and they can be expressed with respect to the pseudo-particle Resolvents
$G_{mn}$ and their self energies $\Sigma_{mn}$ where $m,n$ are
indices representing the eigenstates of the cluster. The NCA equations,
which are used to evaluate the updated resolvent self-energies along
with the cluster Green function $G_{\mu\nu}$ for a given hybridization
function $\Delta_{\mu\nu}$, are \citep{kotliar:865}:

\begin{eqnarray}
\Sigma_{mn}(i\omega) & = & \sum_{m^{\prime}n^{\prime}\mu\nu}F_{\nu}^{mm^{\prime}}\left(F_{\mu}^{n^{\prime}n}\right)^{*}\times\nonumber \\
 &  & \int d\xi f(\xi)\Delta_{\mu\nu}(\xi)G_{m^{\prime}n^{\prime}}\left(i\omega+\xi\right)\nonumber \\
 &  & +\sum_{m^{\prime}n^{\prime}\mu\nu}\left(F_{\nu}^{mm^{\prime}}\right)^{*}F_{\mu}^{n^{\prime}n}\times\nonumber \\
 &  & \int d\xi f(-\xi)\Delta_{\mu\nu}(\xi)G_{m^{\prime}n^{\prime}}\left(i\omega-\xi\right)\label{eq:NCA_SelfEnergy}\\
G_{\mu\nu}(i\omega) & = & -\frac{1}{Q}\sum_{mnm'n'}F_{\mu}^{n'n}\left(F_{\nu}^{m'm}\right)^{*}\times\nonumber \\
 &  & \int\frac{d\xi}{\pi}e^{-\beta\xi}\left[\overline{G}_{m'n'}(\xi)G_{nm}(\xi+i\omega)\right.\nonumber \\
 &  & \left.-G_{m'n'}(\xi-i\omega)\overline{G}_{nm}(\xi)\right]\label{eq:NCA_Cluster_Greens}\\
Q & = & -\int\frac{d\xi}{\pi}e^{-\beta\xi}\sum_{m}\overline{G}_{mm}(\xi)\nonumber \end{eqnarray}
The Greek indices correspond to cluster degrees of freedom (site and
spin), $F_{\nu}^{mm^{\prime}}=\left\langle m\left|c_{\nu}\right|m^{\prime}\right\rangle $
are the matrix elements of the destruction operator and $f(\xi)$
is the Fermi function. The resolvents can be obtained from the self
energy using the Dyson equation,\begin{equation}
G_{mn}(i\omega)=\left(i\omega-E-\lambda-\Sigma\right)_{mn}^{-1}\label{eq:Dyson_Equation}\end{equation}
where $E$ is the diagonal matrix of the clusters eigen-energies and
$\lambda$, an artifact of the slave-boson approach, is chosen for
convenience.

In the paramagnetic, disordered state, both the spin and the irreducible
representation of the geometrical symmetry group are good quantum
numbers and can be used to label the cluster eigenstates,\begin{equation}
\left|m\right\rangle =\left|N,S^{2},S_{z},r,r_{m},E\right\rangle \label{eq:Good_Quantum_Numbers}\end{equation}
where $r$ and $r_{m}$ denote the irreducible representation and
it's row, respectively and the rest follow standard notation. All
but the energy $E$ are good quantum numbers and they cannot be affected
by the bath. Furthermore all resolvents with the same $N,S^{2},r$
but different $S_{z}$ and $r_{m}$ are the equal to one another.

\subsection{Self consistency condition\label{sub:Self-consistency-condition}}

The self consistency condition generates a new hybridization function,
i.e., a new effective environment for the impurity cluster, starting
from a given cluster Green function $G_{\mu\nu}$, and taking into
account the geometry of the lattice and the non-interacting hoping
matrix. Within the CDMFT scheme, the self consistency condition reads

\begin{equation}
\sum_{{\bf K}}(M^{-1}-E(K))^{-1}=(M^{-1}-\Delta-T_{0})^{-1}\label{eq:general_self_cons}\end{equation}
where the matrix $E({\bf K})$ is the Fourier transform of the inter-cluster
hoping matrix $T_{IJ}$, with $I$ and $J$ being indices that label
the clusters inside the superlattice, and ${\bf K}$ is the superlattice
momentum. The cluster cumulant $M_{\alpha\beta}$ can be expressed
in terms of the intra-cluster hoping matrix $T_{0}=T_{II}$ as\begin{equation}
M^{-1}=G^{-1}+\Delta+T_{0}.\label{eq:def_cumulant}\end{equation}

\subsection{Lattice periodization\label{sub:Lattice-periodization}}

Once the cluster quantities such as the cumulant or the self energy
have been obtained, the corresponding lattice quantities need to be
reconstructed \citet{Tudor_CDMFT_2006}. A good estimate of these
lattice quantities can be obtained by averaging over all the possible
ways in which a lattice can be covered with clusters of a given type.
For each type of cluster and each lattice there are a finite number,
$N_{S}$, of different realizations of the superlattice, which are
related to one another by a symmetry operation (rotation, translation
or both). Explicitly, a lattice quantity $X_{latt}({\bf x}_{\alpha}-{\bf x}_{\beta})$,
which may be either the cumulant of the self energy, can be extracted
from the corresponding set of cluster components $X({\bf x}_{\alpha},{\bf x}_{\beta})$
as\begin{equation}
X_{latt}({\bf x}_{\alpha}-{\bf x}_{\beta})=\frac{1}{N_{S}}\sum_{S}X^{SL}(S[{\bf x}_{\alpha}],S[{\bf x}_{\beta}])\label{eq:Periodization_symmetric_average}\end{equation}
where $X^{SL}$ represent the quantity $X$ for a certain reference
superlattice, $S$ is a symmetry operation relating different equivalent
superlattices to the reference superlattice and $S[{\bf x}_{\alpha}]$
is the new position of site ${\bf x}_{\alpha}$ after applying the
symmetry operation. If the positions $S[{\bf x}_{\alpha}]$ and $S[{\bf x}_{\beta}]$
do not belong to the same cell of the reference superlattice $X^{SL}(S[{\bf x}_{\alpha}],S[{\bf x}_{\beta}])$
vanishes, otherwise it is given by the corresponding cluster component.
The momentum dependence is obtained by a simple Fourier transform,\begin{equation}
X_{P}({\bf k})=\sum_{\beta}X_{latt}({\bf x}_{\alpha}-{\bf x}_{\beta})e^{i({\bf x}_{\alpha}-{\bf x}_{\beta})\cdot{\bf k}}\label{eq:Periodization_FT}\end{equation}
where the index $P$ signifies that the quantity $X({\bf k})$ was
obtained by applying the periodization procedure.

In this study, we focus on the cumulant periodization scheme $X\equiv M$,
in which the lattice Green function is given by

\begin{equation}
G(\omega,{\bf k})=\frac{1}{M_{P}^{-1}(\omega,{\bf k})-\epsilon({\bf k})}\label{eq:Lattice_Greens_Function}\end{equation}
We also discuss briefly the implications of using the self energy
periodization scheme. Note that the cluster self energy is related
to the cluster cumulant through the matrix equation

\begin{equation}
\Sigma=\left(\omega+\mu\right){\bf 1}-M^{-1}\label{eq:SigmaFromCumulant}\end{equation}
and a similar scalar equation holds for the corresponding lattice
quantities, $\Sigma({\bf k})=\left(\omega+\mu\right)-M({\bf k})^{-1}$.
In this scheme the lattice Green function is given by

\begin{eqnarray}
G(\omega,{\bf k}) & = & \frac{1}{\omega+\mu-\epsilon({\bf k})-\Sigma_{P}(\omega.{\bf k})}\nonumber \\
 & = & \frac{1}{\left(M^{-1}\right)_{P}(\omega,{\bf k})-\epsilon({\bf k})},\label{eq:Lattice_GreensFunctionFromSigma}\end{eqnarray}
where the index $P$ implies that $M^{-1}$ is periodized. 

In this study we are going to use two types of clusters: a triangular
3-site cluster and a rhombic 4-site cluster. Because of paramagnetism
and also the geometrical symmetry, the triangular cluster has 31 independent
resolvents and 2 independent cluster quantities, whereas the rhombic
cluster has 309 and 5 respectively. Single site and two-site clusters
have also been considered but convergence is possible only at higher
temperatures. The respective superlattices are shown in Fig. \ref{fig:2cluster-superlattice.}. 

For the triangular cluster, there are only two independent components,
a local $X_{0}$ and the nearest neighbor one $X_{1}$. The corresponding
periodization is\[
X_{tri}({\bf k})=X_{0}+2X_{1}a({\bf k})\]
where $a({\bf k})=\frac{1}{3}\sum_{i=1}^{3}\cos k_{i}$ and $k_{1}=k_{x}$,
$k_{2}=-\frac{1}{2}k_{x}+\frac{\sqrt{3}}{2}k_{y}$, $k_{3}=-\frac{1}{2}k_{x}-\frac{\sqrt{3}}{2}k_{y}$.
For the rhombic cluster there are 5 independent components: 2 local
ones $X_{0}$ and $X_{0}^{\prime}$ corresponding to the site with
3 and 2 neighbors inside the cluster respectively, two nearest neighbors
$X_{1}$ and $X_{1}^{\prime}$ corresponding to one of the sides of
the cluster and the diagonal link respectively and one next-to nearest
neighbor component $X_{2}$ along the long diagonal of the cluster.
The periodization is:\[
X_{rho}({\bf k})=\frac{X_{1}+X_{2}^{\prime}}{2}+\left(2X_{1}+\frac{1}{2}X_{1}^{\prime}\right)a({\bf k})+\frac{X_{2}}{2}b({\bf k})\]
where $b({\bf k})=\frac{1}{3}\sum_{i=1}^{3}\cos\left(k_{i}-k_{i+1}\right)$
with $k_{4}=k_{1}$. Since a uniform paramagnetic phase is assumed,
the choice of the cluster is expected to have a relatively small impact
on the physical quantities for regimes characterized by short correlation
lengths.

\begin{figure}
\includegraphics[width=0.8\linewidth]{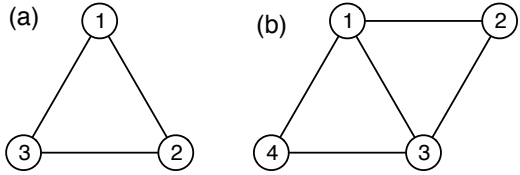}

\caption{\label{fig:Types-of-clusters}Types of clusters used in the simulation:
(a) triangle (b) rhomboid.}

\end{figure}

\begin{figure}
\includegraphics[width=0.5\linewidth]{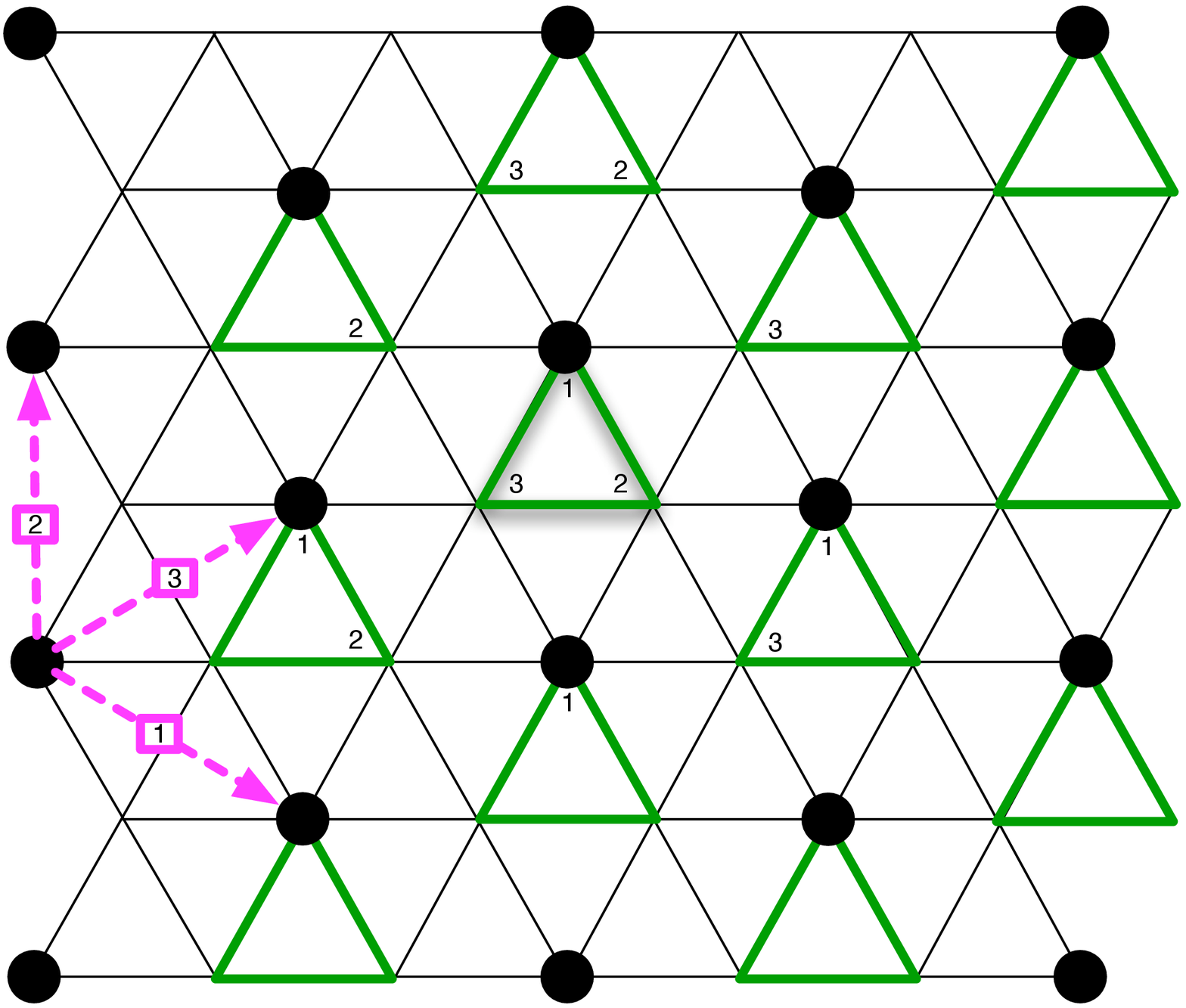}\includegraphics[width=0.5\linewidth]{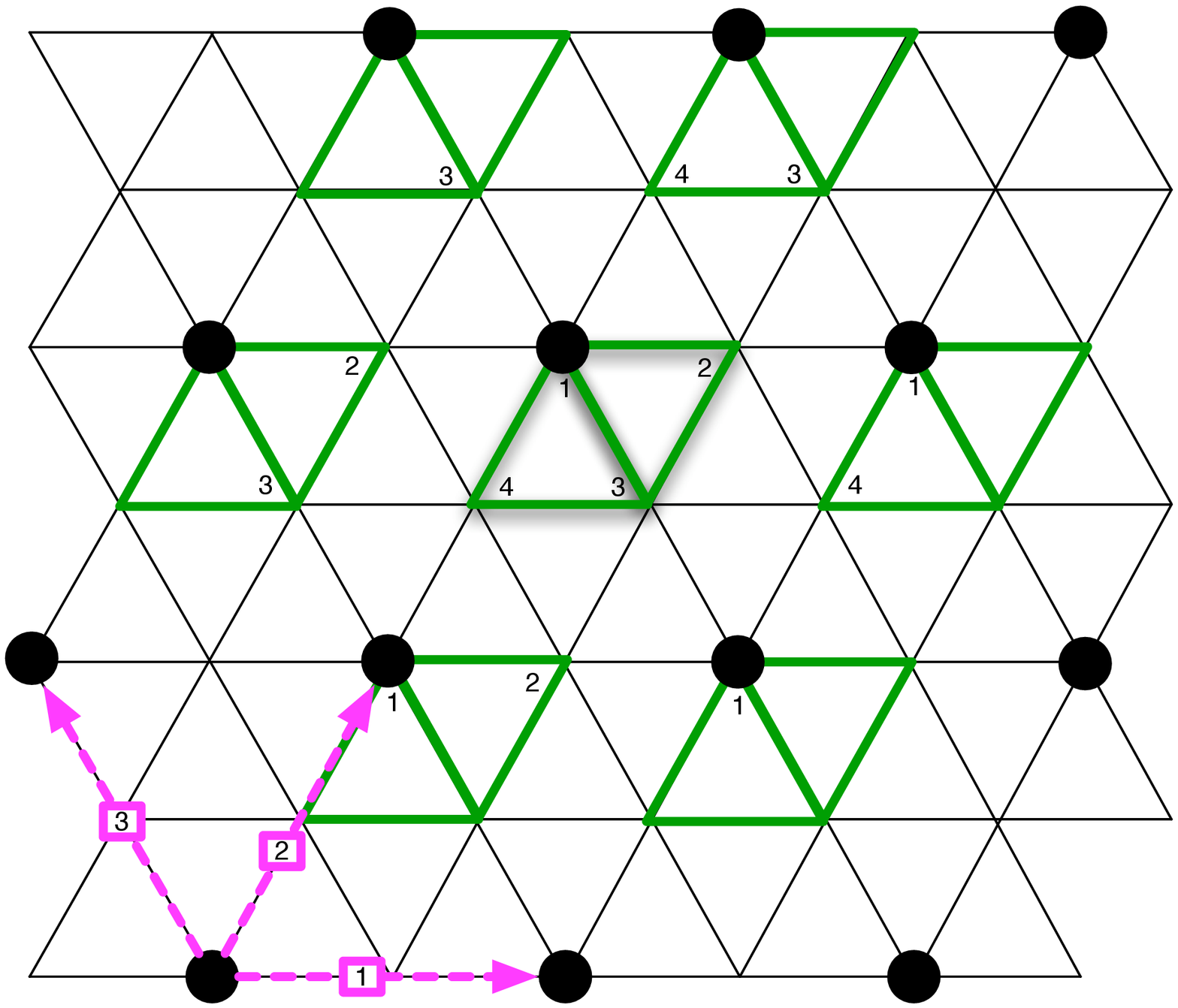}

\caption{\label{fig:2cluster-superlattice.}Triangular (left) and Rhomboid
(right) cluster's superlattice. The black circles represent the superlattice's
sites and the dashed arrows its unit vectors.}

\end{figure}

\begin{figure}
\includegraphics[width=1\columnwidth]{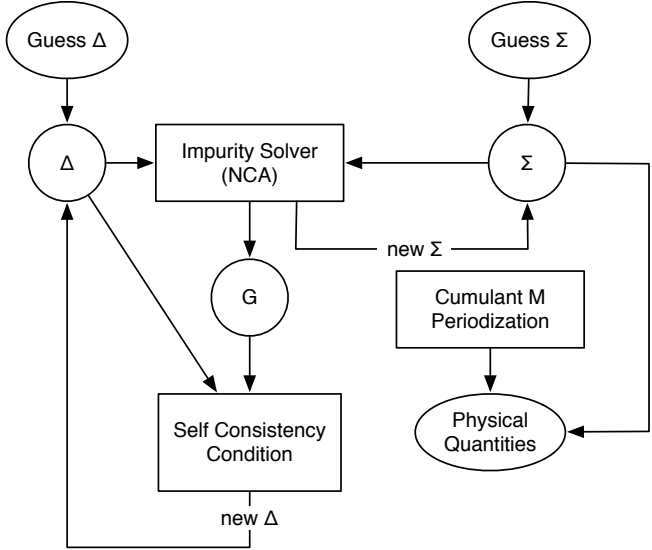}

\caption{Flow diagram for the CDMFT+NCA method. $\Delta$ is the bath function,
$\Sigma$ is the resolvent self energy and $G$ the cluster Green
function and the cumulant is $M=G^{-1}+\Delta+T_{0}$$ $. The flow
starts at the Guesses of $\Delta$ (usually with a Gaussian imaginary
part) and $\Sigma$ (negative imaginary constant). The impurity solver
evaluates a new $\Sigma$ from $\Delta$ and the old $\Sigma$ and
also the cluster Green function. The self consistency condition takes
as its input the cluster Green function and the old $\Delta$ to return
an updated $\Delta$. The impurity solver and self consistency condition
iterate until convergence is reached. Then the cumulant $M$ is evaluated
and periodized. Physical quantities can be obtained from $M$ (for
example, the spectral function) and $\Sigma$. \label{fig:Flow-diagram}}

\end{figure}

\section{Numerical scheme: consistency, optimization and limitations\label{sec:Limitations}}

The key feature that makes a DMFT-type treatment applicable is the
locality of the correlated physics. In infinite dimensions, the correlations
are purely local and can be described by a momentum-independent self-energy.
In finite dimensions, the basic assumption is that the correlations
are short ranged and can be captured by a cluster extension of DMFT.
The size of the cluster that would properly capture the physics is
determined by the range of the relevant correlations and cannot be
known \emph{a priori}. Therefore, consistency checks are a necessary
component of any cluster DMFT treatment. In this section, we show
that for the two-dimensional Hubbard model on a triangular lattice:
A) The self-energy is not a short-range quantity in the Mott insulating
phase and nearby and therefore should not be extracted from the cluster
components. Instead, the renormalized two-point cumulant satisfies
the locality requirement and can be used for re-constructing the lattice
quantities. B) A cluster scheme does not work equally well for all
doping values. In particular, for certain doping levels commensurate
with the cluster size the scheme predicts spurious {}``insulating''
states. We argue that a comparison between results obtained using
clusters of different sizes is crucial. The ultimate consistency criterion
is the invariance of results to an increase of the cluster.

To solve for the cluster quantities and the resolvents self consistently,
we start from an initial guess for the imaginary part of the resolvent
self energies and the hybridization function. The real part was obtained
through the Krammers-Kronig relationships. One possibility is to start
at high temperatures ($T\approx0.3t$) where the method converges
very easily (a constant $\Im\Sigma_{mn}$ and a Gaussian $\Im\Delta_{\mu\nu}$
is enough), and then {}``cool down'' progressively using in every
step the solution of the previous step. Usually a two step process
suffices. Once the initial guess is obtained, the NCA equations, (\ref{eq:NCA_SelfEnergy})
and (\ref{eq:NCA_Cluster_Greens}), along with Eq. (\ref{eq:Dyson_Equation})
are used to update $\Sigma_{mn}$ and evaluate the cluster Green function
$G_{\mu\nu}$. From the latter and the self consistency condition,
Eq. (\ref{eq:general_self_cons}), the hybridization function $\Delta_{\mu\nu}$
is updated. The process iterates until convergence is reached. The
lattice Green function and the corresponding spectral function are
obtained by the periodized cumulant. A flow diagram of the process
is shown in Figure \ref{fig:Flow-diagram}. From the spectral function,
a variety of two-particle properties can be obtained.

\subsection{Cumulant versus self-energy periodization\label{CumSen}}

\begin{figure}
\includegraphics[width=1\columnwidth]{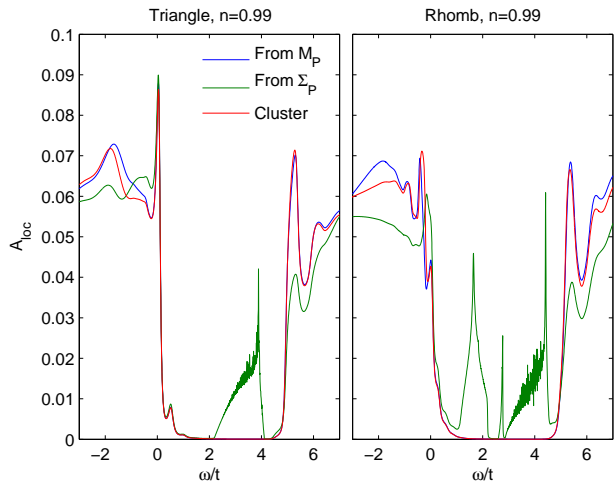}

\caption{Comparison of the local cluster spectral function, $A_{ii}$, with
the one obtained through cumulant $M_{P}$ and self-energy $\Sigma_{P}$
periodization, for the triangular (left) and rhombic cluster (right)
at $U=12t$ and $T=0.1t$. The local cluster spectral function agrees
with that from $M_{P}$. The one obtained from $\Sigma_{P}$ predicts
unphysical states in the middle of the Mott gap.\label{fig:Compare_A_MP_SigP_Rhomb_Triangle}}

\end{figure}

\begin{figure}
\includegraphics[width=1\columnwidth]{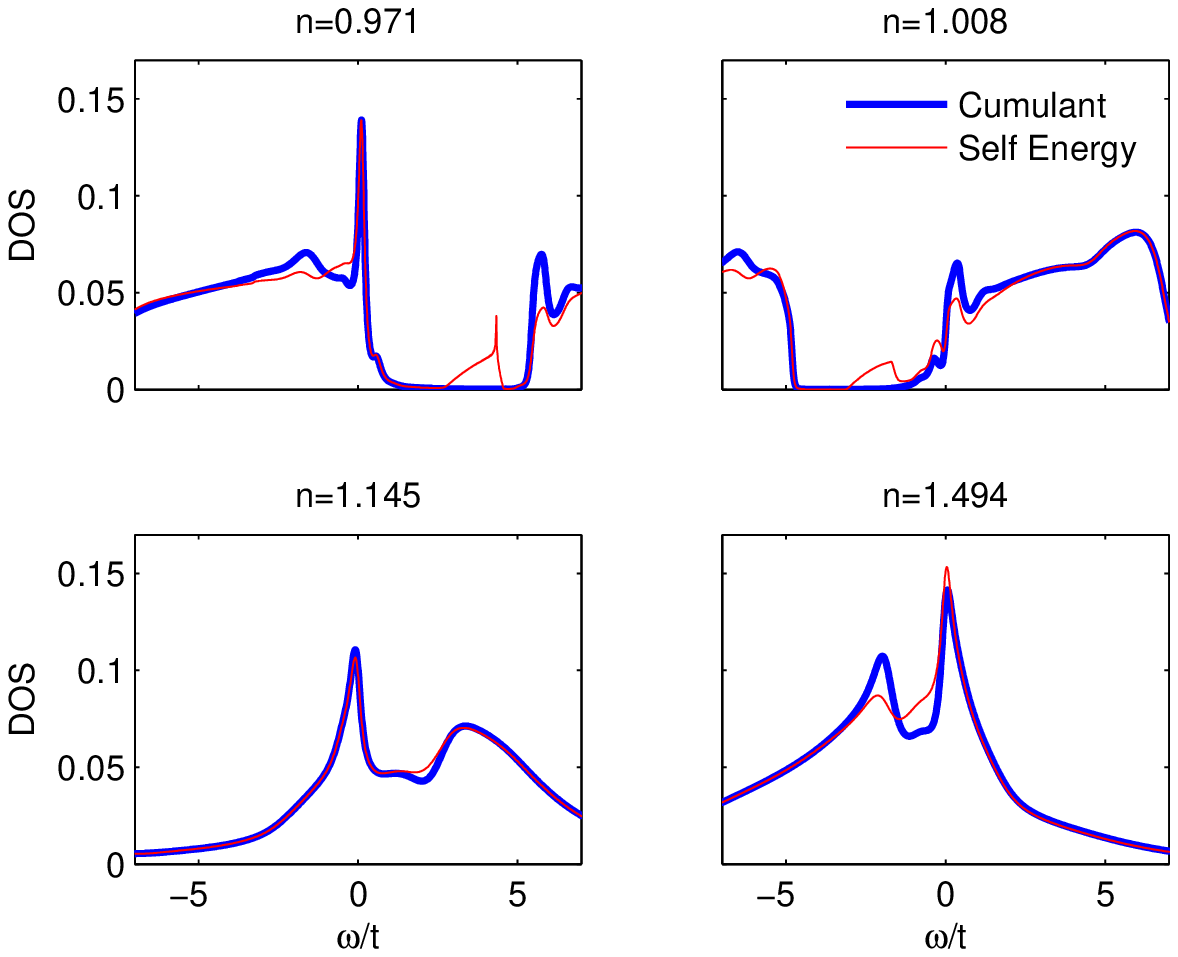}\caption{Comparison of the density of states for the cumulant and self energy
periodization for the triangular cluster and U=12t, T=0.1t.\label{fig:Triangle_Periodization_Comp}}

\end{figure}

\begin{figure}
\includegraphics[width=1\columnwidth]{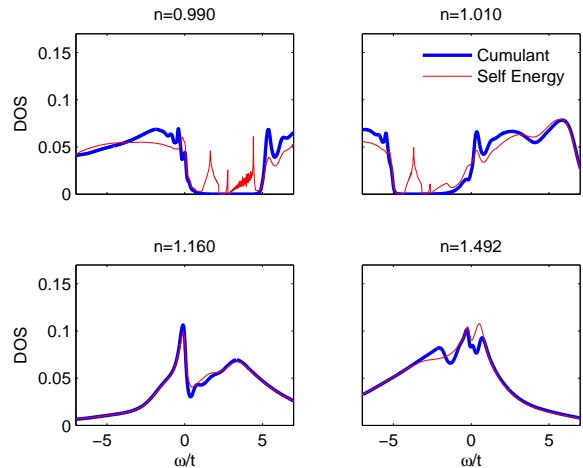}

\caption{Comparison of the density of states for the cumulant and self energy
periodization for rhombic cluster at U=12t, T=0.1t. \label{fig:Rhomb_Periodization_Comp}}

\end{figure}

Our first task is to test the accuracy of the periodization procedure
and identify the quantity most suitable to be used in the periodization
scheme. Note that the implicit physical assumption behind periodizing
a certain quantity $X({\bf x}_{\alpha},{\bf x}_{\beta})$ is the short-range
nature of that quantity. A long-range quantity cannot be properly
approximated in any way using finite range cluster components.

To chose the proper periodization procedure, we compare the density
of states for cumulant and self-energy periodization for both 3 and
4 site clusters. In the vicinity of half-filling, regardless of the
cluster size, the self energy periodization results in states lying
inside the Mott gap, as shown in Fig. \ref{fig:Triangle_Periodization_Comp}
and\ref{fig:Rhomb_Periodization_Comp}. These states are clearly unphysical,
as demonstrated by the comparison with the local cluster spectral
function, which shows a well-defined, clean Mott gap, as shown in
Fig. \ref{fig:Compare_A_MP_SigP_Rhomb_Triangle}. On the other hand,
these mid-gap states are absent in the cumulant periodization procedure.
This signals that the cumulant is a short-range quantity in the vicinity
of the Mott transition, while the self-energy is not, but contains
long-range components that cannot be captured with a small size cluster
\citet{Stanescu_Fermi_Arcs}.

In contrast, at large dopings in the Fermi liquid phase, where both
the self-energy and the cumulant are short-range quantities, the two
methods agree. We conclude that the self-energy periodization is appropriate
away from the Mott transition, while the cumulant scheme gives consistent
results in a wide range of dopings. In the present study we will use
the cumulant method regardless of filling. The reason for the failure
of the self-energy periodization method is the presence in the half-filled
regime of self-energy divergences \citep{Tudor_CDMFT_2006,stanescu_zeros}
at $\omega=0$ and low temperatures. This divergence of the self energy
at half-filling is intimately linked to the Mott gap.

\subsection{Commensurate Insulators\label{sub:Commensurate-Insulators}}

\begin{figure}
\includegraphics[width=0.5\columnwidth]{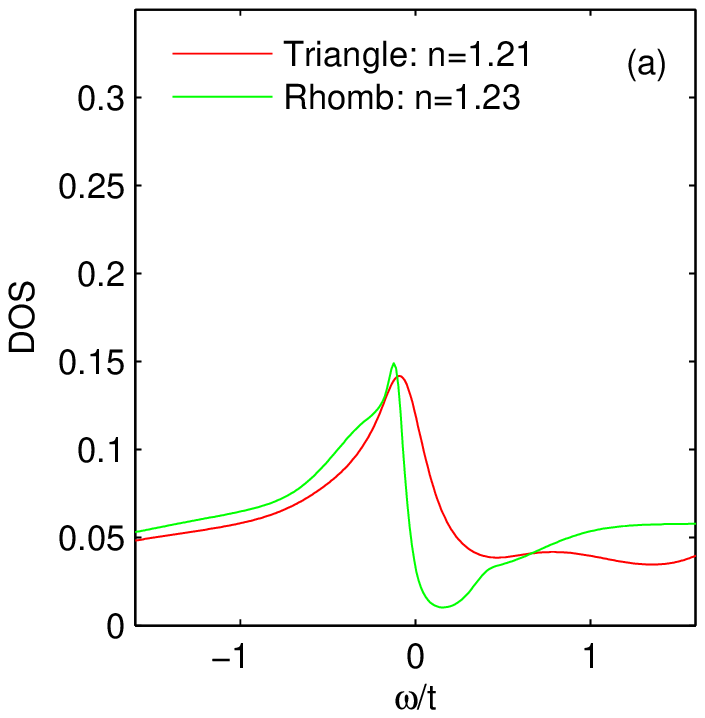}\includegraphics[width=0.5\columnwidth]{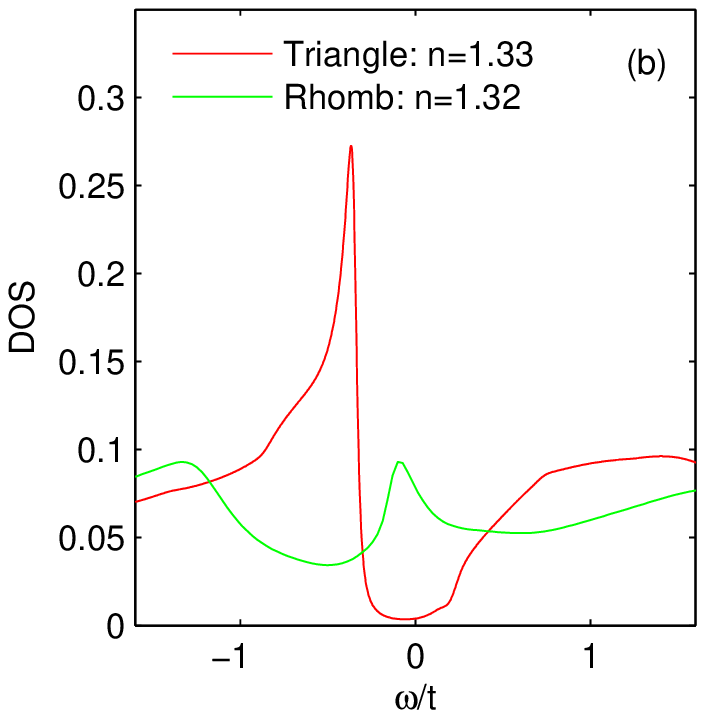}

\caption{The density of states for both clusters for (a) $n=1.22$ and (b)
$n=1.33$\label{fig:DOS_compare_commensurate_ins}}

\end{figure}

Away from half filling, in the electron doped regime, there is a substantial
discrepancy between the two clusters. More explicitly, our method
predicts that around $n\approx1.33$ for the triangular and $n\approx1.25$
for the rhombic cluster, the system becomes an insulator as evidenced
by the gap in the density of states in Fig. \ref{fig:DOS_compare_commensurate_ins}
(a) and (b) respectively.

\begin{figure}
\includegraphics[width=0.5\columnwidth]{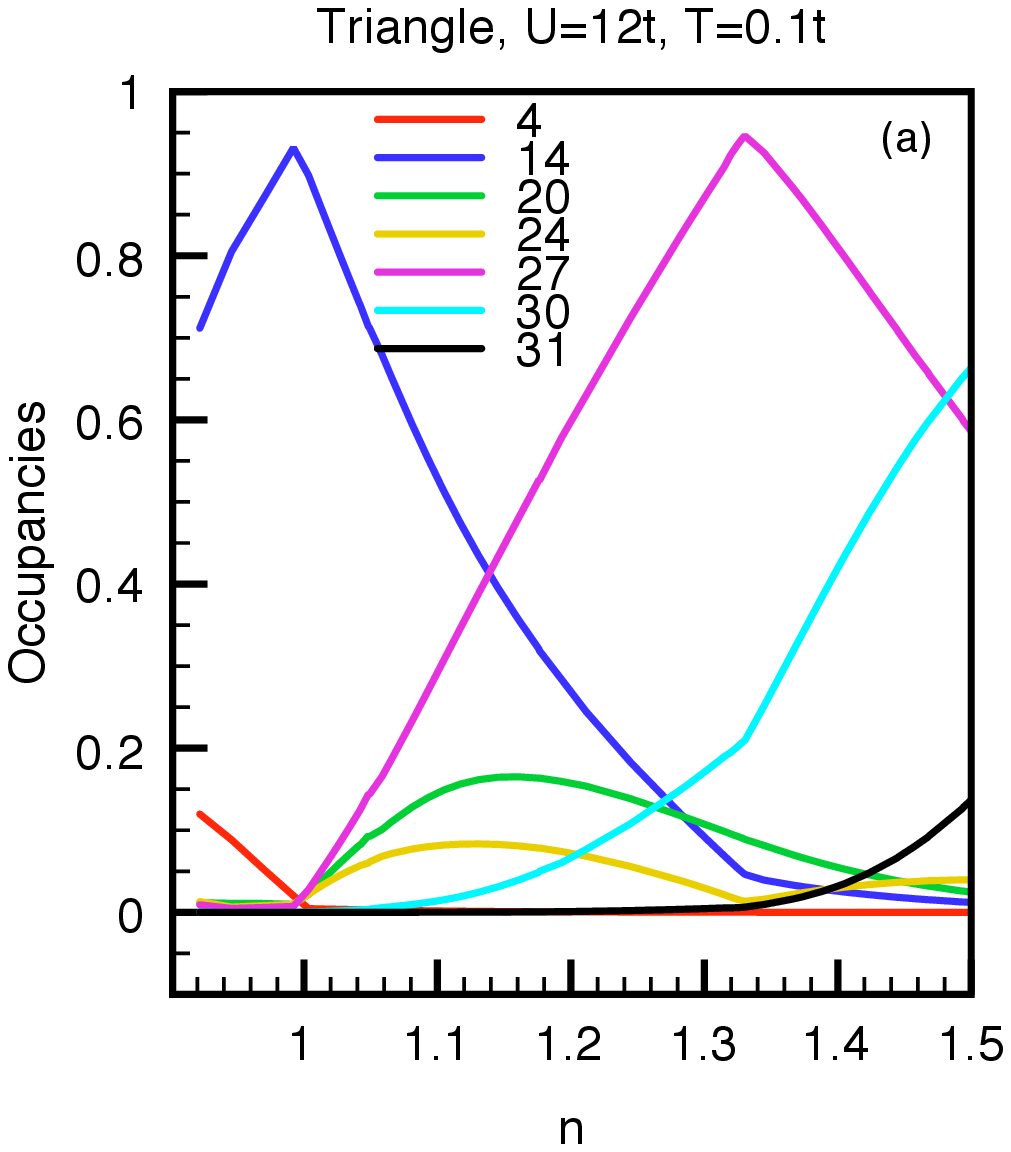}\includegraphics[width=0.5\columnwidth]{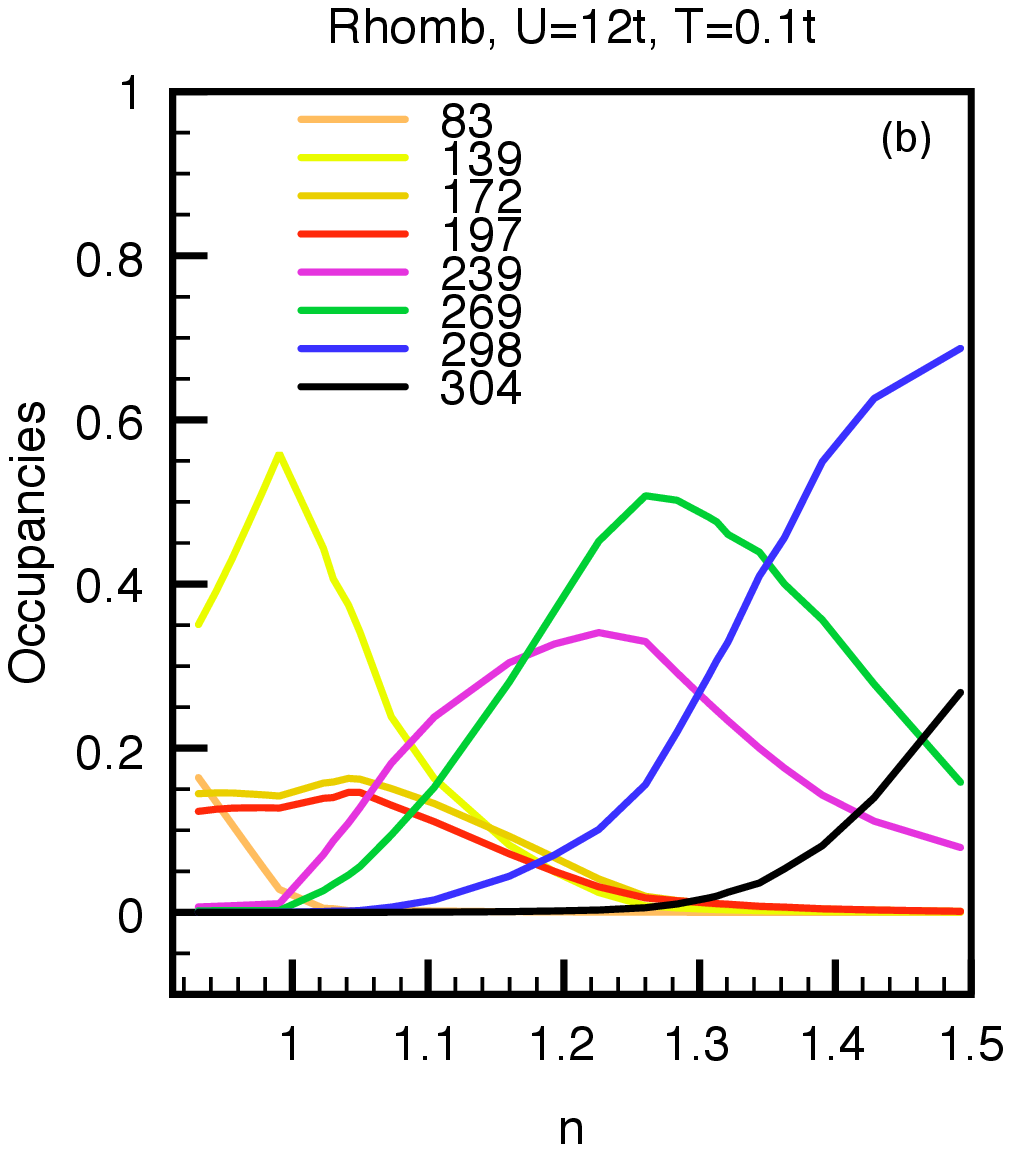}

\caption{The occupancies versus filling for the dominant resolvents in the
triangular (a) and rhombic (b) cluster.\label{fig:Overlap-vs-Filling}}

\end{figure}

To better understand the commensurate insulator phases, we can analyze
the resolvents within NCA and evaluate the contribution of each cluster
state. It is therefore possible to determine which are the dominant
channels through which the cluster interacts with the bath. The relevant
quantity is the partial occupancy of a particular cluster state\begin{equation}
\left\langle n{}_{mm}\right\rangle =-\frac{1}{Q}\int\frac{d\xi}{\pi}e^{-\beta\xi}n_{mm}\overline{G}_{mm}(\xi).\label{eq:overlap}\end{equation}
Small occupancy signifies that the corresponding resolvent contributes
insignificantly to the cluster spectral function and consequently
to the hybridization function. Therefore, the resolvents with small
occupancy can be ignored, whereas the ones with large overlap have
the dominant contribution to the dynamics of the system. For the triangular
cluster it turns that there are only 7 resolvents with an overlap
larger than $0.01$ whereas the rest have occupancy less than $0.0025$.
A similar analysis can be performed for the rhombic cluster. As expected
there are more resolvents with substantial occupancy. In Fig. \ref{fig:Overlap-vs-Filling}
only the dominant ones are plotted which have occupancy more than
0.1. Their quantum numbers are shown in Table \ref{tab:RhombDominantQN}.
In all cases, we observe that when the dominant resolvents have fillings
$i/N_{c}$ close to the lattice filling, their partial occupancy peaks
and transitions from and to them become rare, which gives rise to
an insulating state. In the triangular lattice this has a consequence
when $n=4/3$, because in this case there is only one dominant resolvent.
As a result, there is no appreciable overlap with any electronic states,
leading thereby to a gap in the spectrum. In the rhombic cluster,
around the critical value $n=5/4$ there are two dominant resolvents
which means that even though the transitions are limited, a gap still
persists, though it is not as pronounced as in the three-site cluster.
We have also obtained commensurate insulating states for fillings
$n=2/3$ for the triangular and $n=3/4$ for the rhombic cluster,
but in this regime convergence is not reached at low temperature.
The presence of such insulating states ultimately points to a limitation
of the finite cluster approach to the Hubbard model.

We emphasize that the appearance of the fictitious commensurate insulating
states is a consequence of using small clusters in the numerical calculations.
The use of these small clusters is dictated by practical reasons.
However, the interpretation of the results requires special care.
Ultimately, the consistency of any result should be confirmed by its
invariance to cluster size variations. Such an example is Mott insulating
phase at half filling, $n=3/3=4/4$. The existence and the properties
of this state can be obtained consistently using various cluster sizes.
Moreover, as the on-site interaction U is reduced, a transition to
a metallic state is observed, with some small cluster size dependence
of the critical parameters. In all the present implementations of
the CDMFT cluster methods for finite dimensions higher than one, there
is always a difficulty related to the fact that most or all of the
cluster sites lie on the cluster boundaries. However, since large
cluster computations are impractical, one has to rely on a careful
interpretation of the small cluster results. We suggest that the ultimate
criterion for the consistency of small cluster calculations is the
independence of the results to the cluster size.

\begin{table}
\begin{tabular}{|r|r|r|r|r|r|r|}
\hline 
Index & $N$ & $S^{2}$ & $S_{z}$ & $r$ & $r_{m}$ & $E$\tabularnewline
\hline
\hline 
4 & 2 & 0 & 0 & 1 & 1 & -2.55\tabularnewline
\hline 
14 & 3 & 1/2 & 1/2 & 3 & 1 & -0.50\tabularnewline
\hline 
20 & 3 & 3/2 & 3/2 & 2 & 1 & 0.00\tabularnewline
\hline 
24 & 4 & 0 & 0 & 1 & 1 & 10.85\tabularnewline
\hline 
27 & 4 & 1 & 1 & 2 & 1 & 10.00\tabularnewline
\hline 
30 & 5 & 1/2 & 1/2 & 3 & 1 & 23.00\tabularnewline
\hline 
31 & 6 & 0 & 0 & 1 & 1 & 36.00\tabularnewline
\hline
\end{tabular}

\caption{The quantum numbers of the dominant resolvents for the triangular
cluster for U=12t, T=0.1t.\label{tab:TriangDominantQN}}

\end{table}

\begin{table}
\begin{tabular}{|r|r|r|r|r|r|r|}
\hline 
Index & $N$ & $S^{2}$ & $S_{z}$ & $r$ & $r_{m}$ & $E$\tabularnewline
\hline
\hline 
83 & 3 & 1/2 & 1/2 & 4 & 1 & -3.07\tabularnewline
\hline 
139 & 4 & 0 & 0 & 2 & 1 & -0.94\tabularnewline
\hline 
1712 & 4 & 1 & 1 & 2 & 1 & -0.67\tabularnewline
\hline 
197 & 4 & 1 & 1 & 4 & 1 & -0.65\tabularnewline
\hline 
239 & 5 & 1/2 & 1/2 & 3 & 1 & 9.39\tabularnewline
\hline 
269 & 5 & 3/2 & 3/2 & 2 & 1 & 9.44\tabularnewline
\hline 
297 & 6 & 1 & 1 & 3 & 1 & 21.44\tabularnewline
\hline 
304 & 7 & 1/2 & 1/2 & 1 & 1 & 34.44\tabularnewline
\hline
\end{tabular}

\caption{The quantum numbers of the dominant resolvents for the rhombic cluster
for U=12t, T=0.1t.\label{tab:RhombDominantQN}}

\end{table}

\section{Results\label{sec:Results}}

In this section we present our main results. In subsection \ref{sub:Dispersionless-low-energy},
we show that the two-dimensional Hubbard model on a triangular lattice
is a strongly correlated system by demonstrating that the low-energy
physics is controlled by a weakly dispersing band with spectral weight
that can be transferred over large energy scales. Subsection \ref{sub:Mott-transition}
is devoted to the characterization of the interaction-controlled Mott
metal-insulator transition at half filling. In contrast to the infinite
dimensional case when a coherence peak develops inside the Mott gap,
we find that the insulator-metal transition is characterized by the
complete collapse of the Mott gap followed by the appearance of a
small peak in the density of states. Finally, in subsection \ref{sub:Pseudogap}
we investigate the system at low dopings and discuss a new perspective
on pseudogap physics. We argue that the pseudogap should not be simply
identified by the depletion of the density of states at the chemical
potential, but rather by the change in the location of the low-energy
modes in momentum space as compared with the non-interacting system.
According to this picture, the pseudogap phase is essentially characterized
by the re-constructed Fermi surface consisting of small pockets that
vanish in the zero doping limit.

\subsection{Dispersionless low energy band\label{sub:Dispersionless-low-energy}}

\begin{figure}
\includegraphics[width=1\columnwidth]{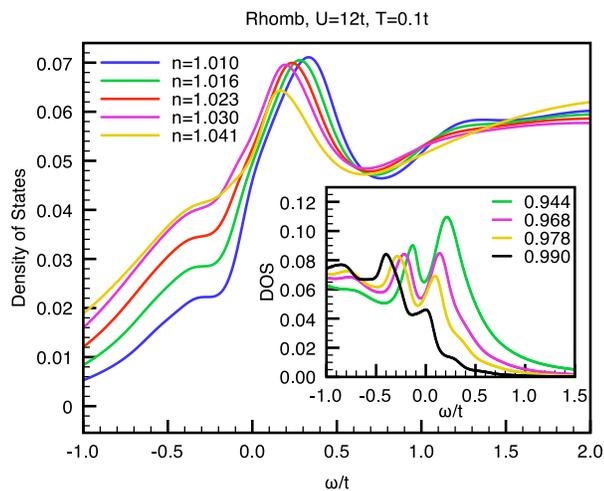}\caption{The evolution of the density of states as a function of doping in
the upper and lower (inset) Hubbard band.\label{fig:RhombDOSEvoHF}}

\end{figure}

\begin{figure}
\includegraphics[width=1\columnwidth]{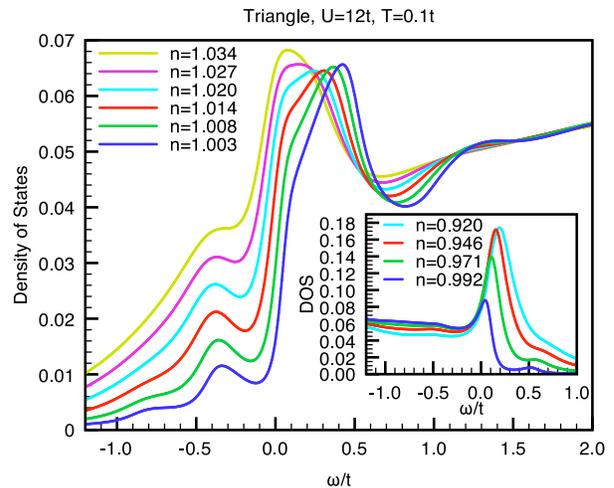}

\caption{Evolution of the density of states for the triangular cluster around
half filling for $T=0.1t$ and $U=12t$ in the electron doped regime
(upper Hubbard band). There is a pseudogap feature which does not
align with the chemical potential $(\omega=0)$ and which disappears
for a doping of $n=3.5\%$. The inset shows the density of states
in the hole doped regime (lower Hubbard band), which exhibits no pseudogap
feature.\label{fig:DOS_TRIANGLE_HF}}

\end{figure}

\begin{figure}
\includegraphics[width=1\columnwidth]{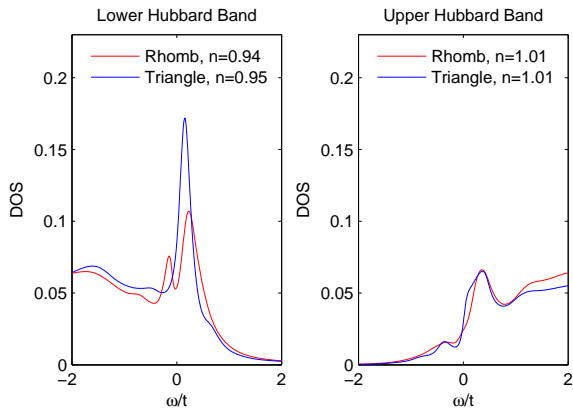}

\caption{Comparison of the density of states between the rhombic and the triangular
cluster in the upper and lower Hubbard band. There is good agreement
in the electron doped regime. In the hole doped regime there is a
dispersionless band which the rhombic clusters resolves as being split.
\label{fig:DOS_two_clusters_comp}}

\end{figure}

\begin{figure}
\includegraphics[width=1\columnwidth]{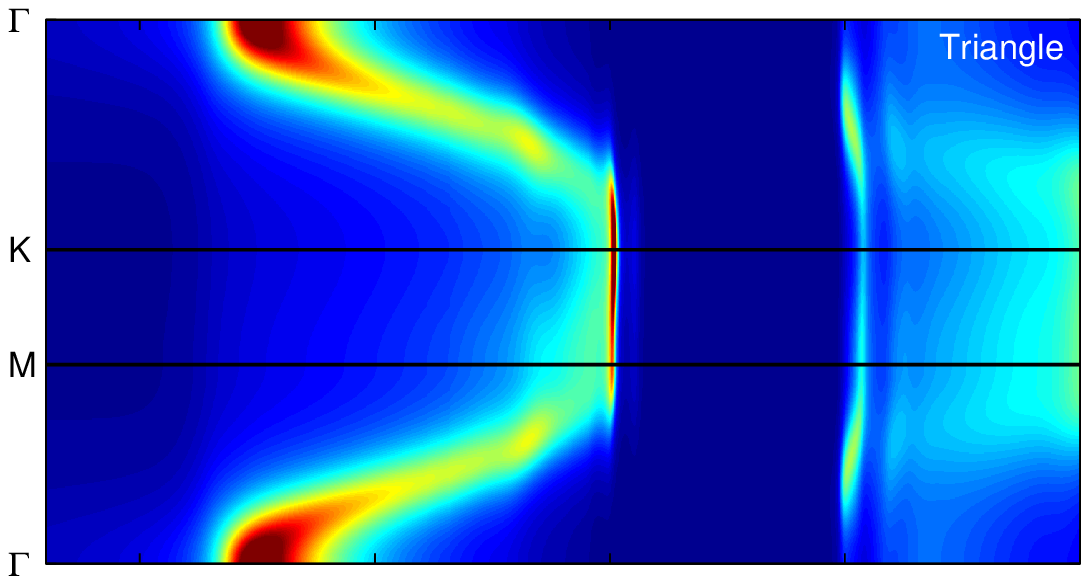}

\includegraphics[width=1\columnwidth]{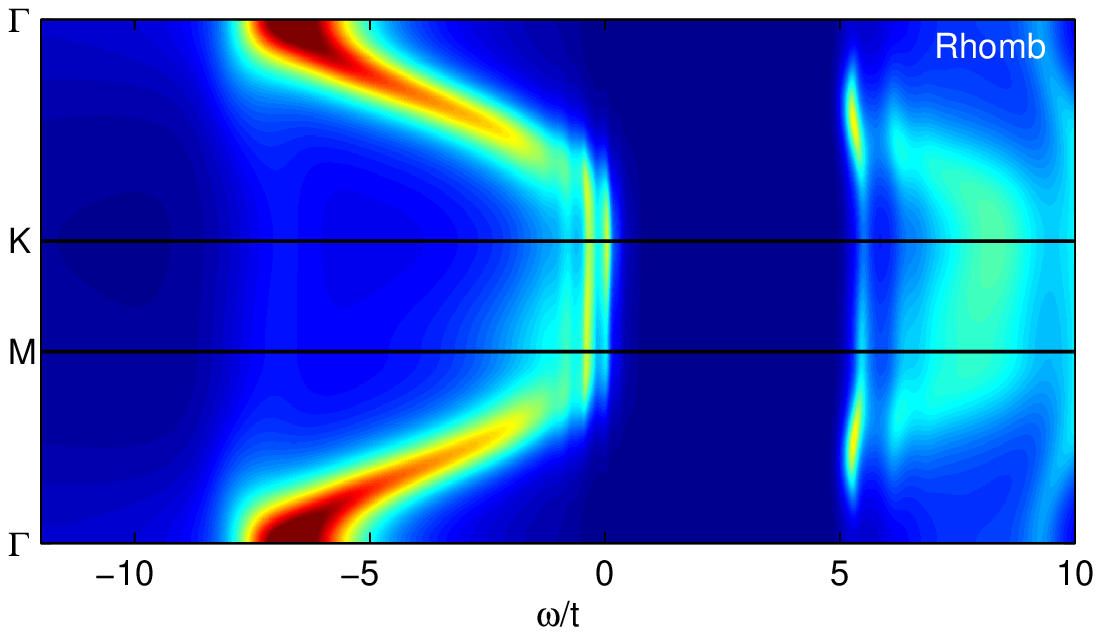}

\caption{The spectral function as a function of $k$ and $\omega$ for $U=12t$,
$T=0.1t$ and $n=0.99$. Upper panel: triangular cluster. lower panel:
rhombic cluster. Both methods give a dispersionless band, but in the
rhombic cluster it is resolved in two.\label{fig:Spectral_nearHF_electron}}

\end{figure}

The main figure and the inset in Fig. \ref{fig:RhombDOSEvoHF} and
Fig. \ref{fig:DOS_TRIANGLE_HF} reveal a lack of particle-hole symmetry
for electron and hole doping. This is expected as a triangular lattice
does not preserve this symmetry. While the asymmetry persists regardless
of the cluster size, the details differ. Of particular interest is
the presence of a dispersionless sub-band residing near the top of
the lower Hubbard band upon hole doping. The occurrence of such a
band is inconsistent with Fermi liquid behavior: the chemical potential
crosses the band in an extended area instead of at a well defined
curve. This band occurs in both the triangular and the rhombic cluster,
but in the latter it appears split. This splitting, shown in Fig.
\ref{fig:DOS_two_clusters_comp}, but also in the density of states
as shown in Fig. \ref{fig:Spectral_nearHF_electron}, may be due to
the higher resolution gained by the using the rhombic cluster.

\begin{figure}
\includegraphics[width=1\columnwidth]{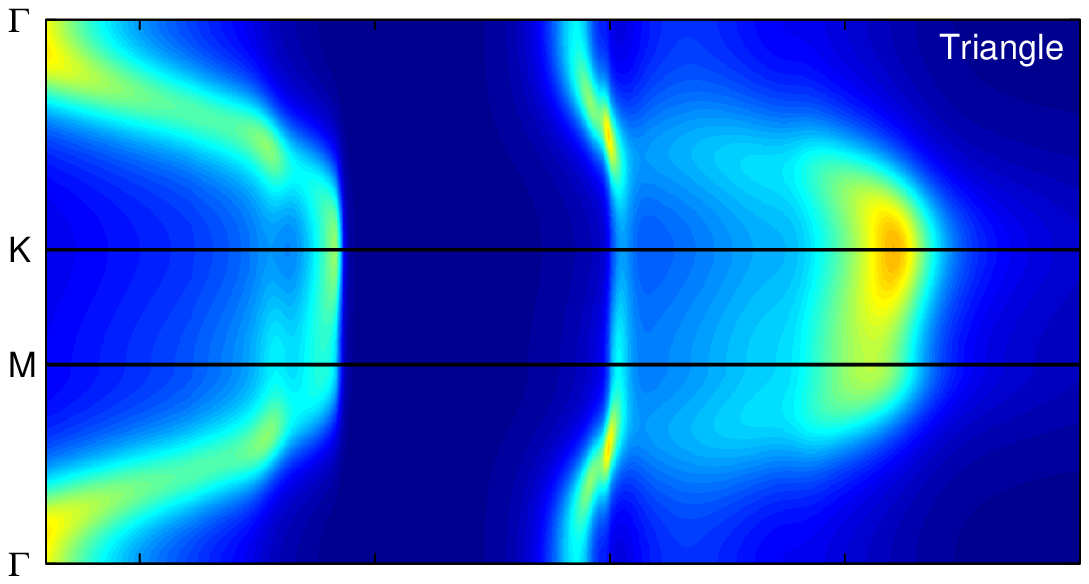}

\includegraphics[width=1\columnwidth]{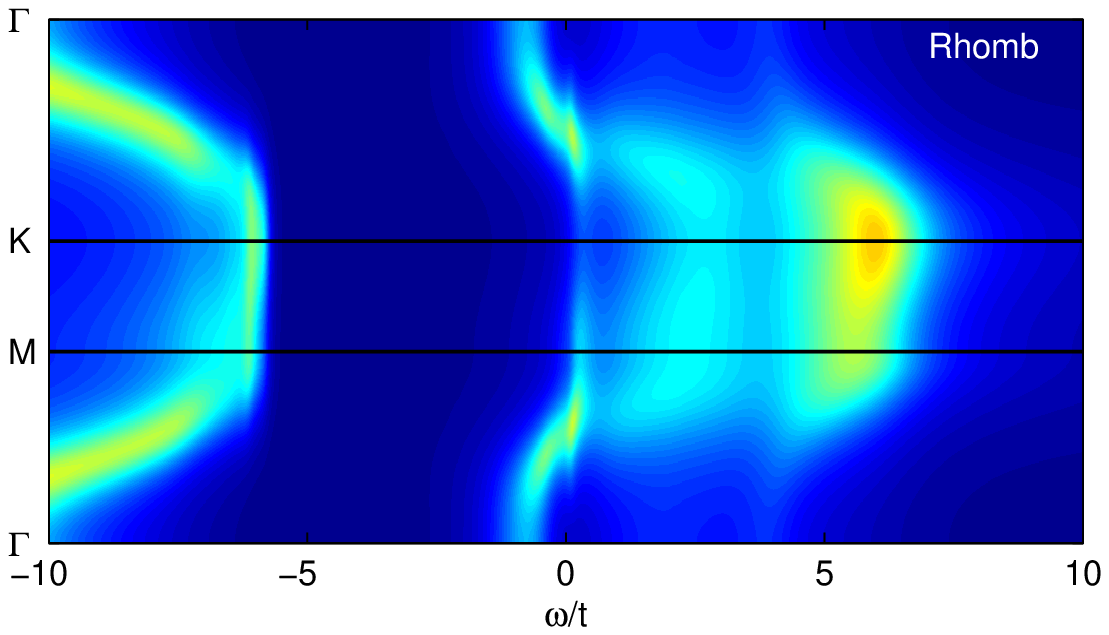}

\caption{Spectral function for $U=12t$, $T=0.1t$ and $n=1.04$ for the triangular
(upper panel) and rhombic cluster (lower panel). The chemical potentials
crosses a narrow band and the behavior is metallic. The dispersionless
sub-band at the top of the lower Hubbard band persists, but with less
spectral weight and unsplit.\label{fig:Spectral_nearHF_hole}.}

\end{figure}

A shadow of this band persists unsplit and with less spectral weight
even in the electron doped regime where the system is a normal metal
as shown in Fig. \ref{fig:Spectral_nearHF_electron}. Therefore the
splitting is due to the fact that it crosses the chemical potential.
Consequently, in the triangular lattice, particle hole asymmetry gives
rise to a non-Fermi liquid behavior for hole doping and metallic behavior
in the electron doped side. This band structure can be compared with
experimental results \citet{qian_Fulld-band_dispersion}. A full comparison
is not possible because only one band, namely the $a_{1g}$ is taken
into account and the $e_{g^{\prime}}$ is ignored. However the experiment
shows the existence of an almost flat band with energy $-0.6eV$.
The present calculation is evidence that this band may emerge purely
because of strong correlations. Further evidence that this band arises
from purely strong electron correlations comes from the fact that
it is absent in local density approximation (LDA) and linear augmented
plane wave (LAPW) calculations \citet[Fig. 3 of][]{qian_Fulld-band_dispersion}.

\subsection{Mott transition\label{sub:Mott-transition}}

\begin{figure}
\includegraphics[width=1\columnwidth]{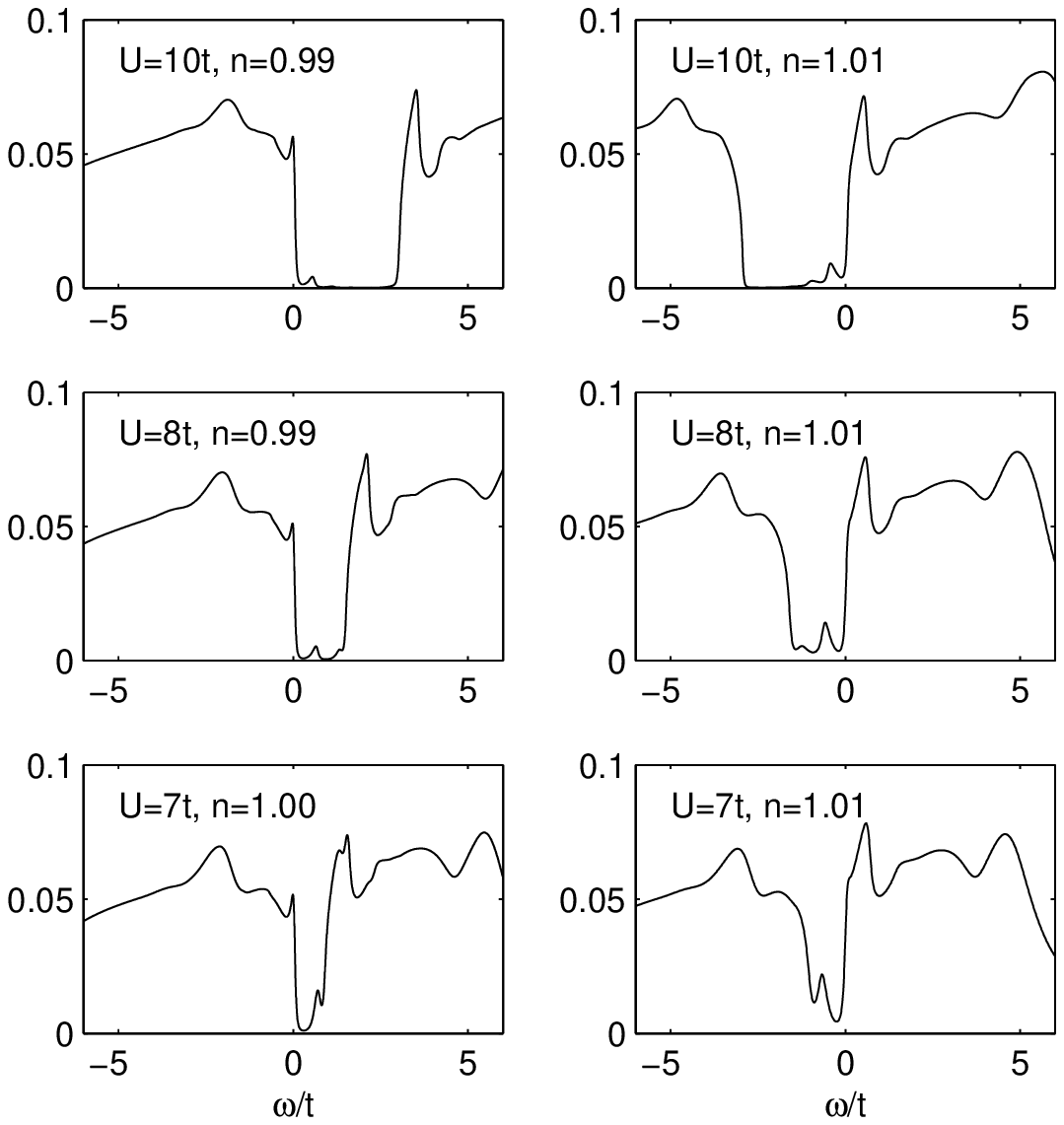}\caption{The density of states in the vicinity of half filling for Triangular
cluster. The gap closes with U.\label{fig:DOSvsU}}

\end{figure}

\begin{figure}
\includegraphics[width=1\columnwidth]{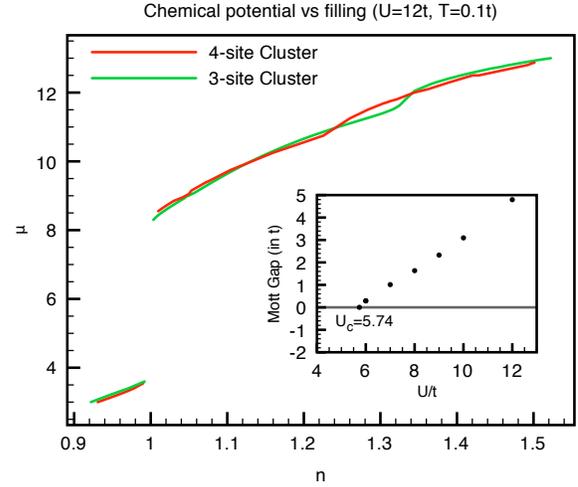}

\caption{The chemical potential as a function of filling for the 3 and 4 site
cluster at $U=12t$ and $T=0.1t$. There is agreement anywhere apart
from the commensurate fillings $n=4/3$ and $n=5/4$ respectively.
Inset: the evolution of the Mott gap as a function of filling for
the triangular cluster. The gap is linear in U and vanishes at $U_{c}=5.74t$.
\label{fig:U12_Mu_vs_Fill}}

\end{figure}

\begin{figure}
\includegraphics[width=0.5\columnwidth]{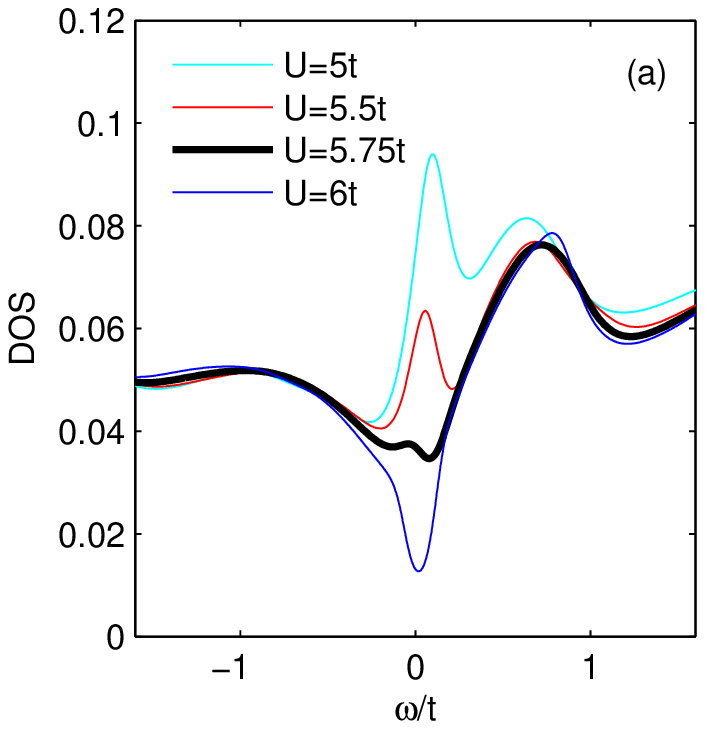}\includegraphics[width=0.5\columnwidth]{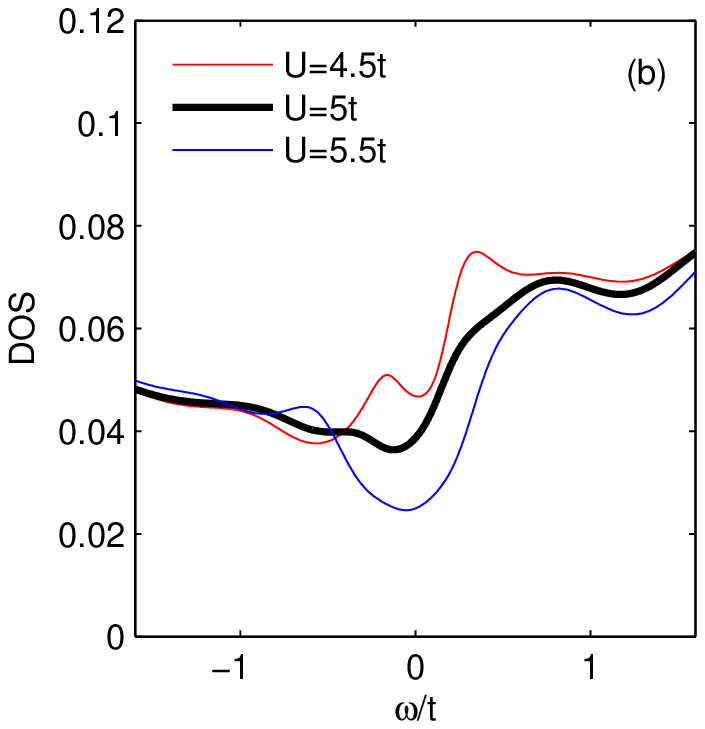}

\caption{(a) density of states around the Mott transition for the triangular
cluster at $n=1$ $T=0.1t$. As $U$ decreases, the Drude peak around
the chemical potential gradually looses weight until it turns turns
into a gap. The transition occurs at some $U_{c}\approx5.75t$. (b):
density of states around the Mott transition for the rhombic cluster
at $T=0.12t$ and $n=1$ and $U_{c}\approx4.5t$. \label{fig:Mott-transition}}

\end{figure}

\begin{figure}
\includegraphics[width=1\columnwidth]{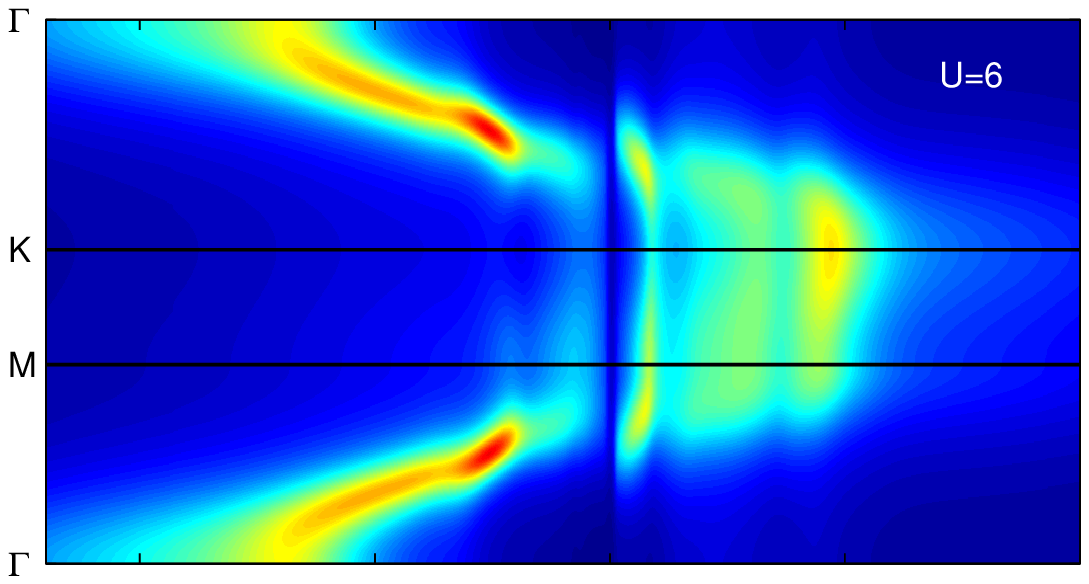}

\includegraphics[width=1\columnwidth]{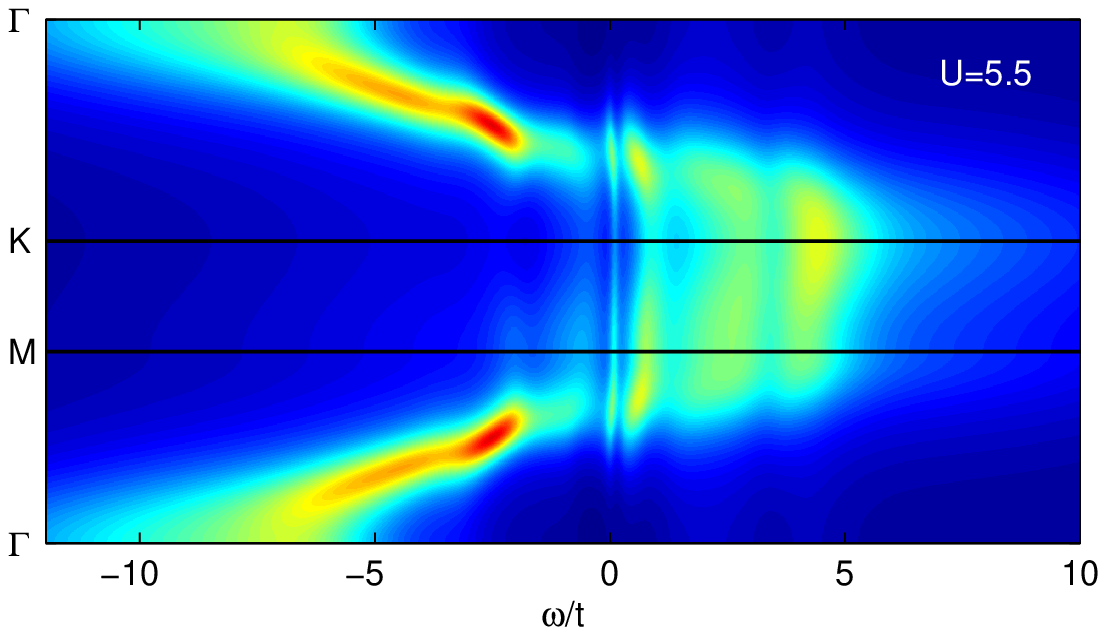}

\caption{The spectral function at half filling and $T=0.1t$ as a function
of $\omega$ and $k$ along the $\Gamma\rightarrow K\rightarrow M\rightarrow\Gamma$
path for $U=6t$ (upper panel) and $U=5.5t$ (lower panel) as obtained
from the triangular cluster. The spectral weight mostly from energy
around $4.5t$ is transferred to the chemical during the transition.\label{fig:Spec_AccrossMottTransition}}

\end{figure}

The density of states for high enough $U$ exhibits an interaction-induced
gap around half filling, as shown in Fig.\ref{fig:DOSvsU}. Clearly
shown in Fig. \ref{fig:DOSvsU} is the closing of the gap between
the lower and upper Hubbard bands as the on-site interaction decreases,
indicative of a Mott transition. To pinpoint the precise location
of the Mott transition, we estimate the Mott gap by the discontinuity,
$\Delta\mu$, in the chemical potential on either side of half-filling.
Fig. \ref{fig:U12_Mu_vs_Fill} displays a typical calculation of the
chemical potential as a function of the filling for both 3 and 4-site
clusters. Because of thermal broadening, this procedure would underestimate
$\Delta\mu$, which cannot discriminate below energy scales of the
order of $kT$. For both cluster sizes, the results are consistent
yielding a gap of $\Delta\mu\approx5t$ for $U=12t$. As the inset
demonstrates, the discontinuity in the chemical potential across half-filling
vanishes at $U_{c}\approx5.7t$. However, $\Delta\mu$ provides only
a rough estimate of the critical $U$ because the precise magnitude
of the gap is obscured as shown in Fig. \ref{fig:U12_Mu_vs_Fill}.
To probe the transition more directly, we plot the density of states
for different values of $U$ around the estimate obtained from the
chemical potential analysis. Fig. \ref{fig:Mott-transition} displays
clearly that for $U\geq U_{c}$, with $U_{c}\approx5.75t$ and $U_{c}\approx4.75t$
for the triangle and rhombic cluster respectively, the system is an
insulator, whereas for $U<U_{c}$ a Drude peak emerges at the chemical
potential. Note that the density of states near the chemical potential
remains unchanged (relative to its value for $U>U_{c}$) although
the density of states at the chemical potential develops a non-zero
value as a transition is made to a metallic state. Consequently, the
states that fill in the Mott gap and give rise to the coherence peak
arise from spectral weight transfer from high energy, the essence
of Mottness. This is illustrated in Fig. \ref{fig:Spec_AccrossMottTransition}
which compares the spectral function right before and after the transition.
Such a redistribution of spectral weight far from the chemical potential
near the Mott transition has been recently observed in the manganites
\citep{Cooper_Sawatzky_Manganites}.

The operative mechanism for the Mott transition in the three and four-site
cluster analysis stands in contrast to the scenario predicted by DMFT
\citep{GeorgesReview}. In this scenario a coherent peak of constant
height exists at the chemical potential, which successively narrows
as $U$ increases to $U_{c}$. For $U>U_{c}$, the peak vanishes and
the upper and lower Hubbard bands become well separated. CDMFT in
the plaquette offers a different scenario \citep{zhang_imada_pseudogap}:
first a pseudogap opens at the chemical potential which smoothly grows
to form a full Mott gap at the critical $U$. In the triangle, there
is no formation of a pseudogap. Instead the coherence peak at the
chemical potential loses weight $Z$ as $U$ increases and is smoothly
replaced by a gap which broadens as the two bands separate.

\subsection{The pseudogap\label{sub:Pseudogap}}

\begin{figure}
\includegraphics[width=1\columnwidth]{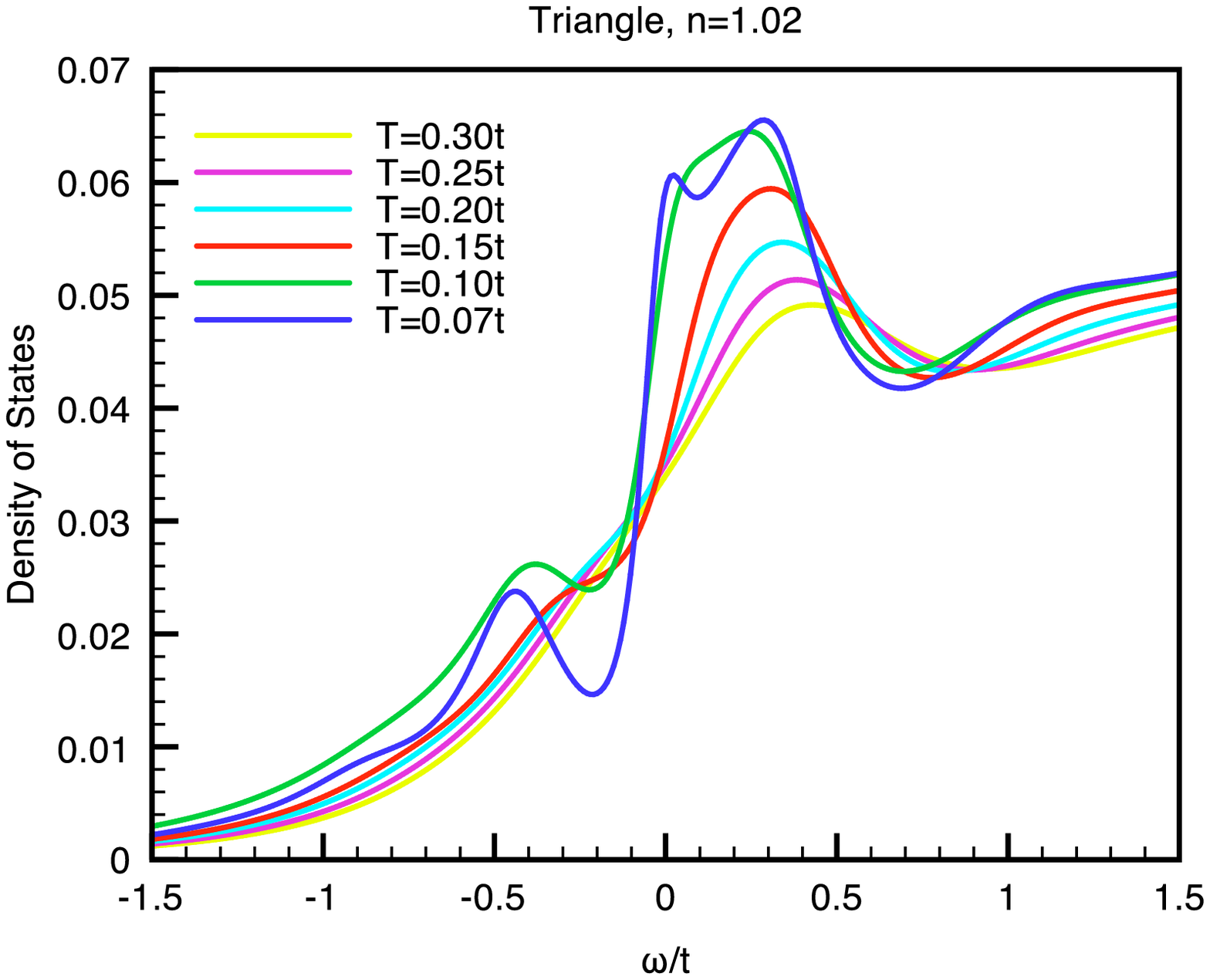}

\caption{Evolution of the DOS around the chemical potential for $n=1.02$ using
the triangular cluster. The pseudogap feature is dispersed as the
temperature increases.\label{fig:DOS_HF_TEMP}}

\end{figure}

In light of the physics in the cuprates, one of the main questions
that needs to be addressed is whether or not a Mott system on a triangular
lattice exhibits a pseudogap. We focus here on the single-particle
density of states as a function of filling. In the triangular lattice,
there is only indirect evidence from a boson analysis of the optical
conductivity for a pseudogap in the paramagnetic phase ($n<1.5$)
\citep{wu_Pseudogap_IR}, \citep{shimojima-pseodogap} which disappears
in the strange metal phase ($n>1.5$). 

Our results for the single-particle density of states on either side
of half-filling are summarized in Figs. \ref{fig:RhombDOSEvoHF} and
\ref{fig:DOS_TRIANGLE_HF}. Although a dip-like features exists for
both the 3 and 4-site clusters, it is displaced from the chemical
potential. For higher dopings, the density of states is smooth in
the vicinity of the chemical potential. Consequently, we find an absence
of a pseudogap near half-filling on a triangular lattice. This result
is consistent with the Quantum Monte Carlo calculations \citet{kyung_CDMFT_Triangle}
as the impurity solver coupled with CDMFT on the triangular lattice.
We traced the origin of the dip-like feature to an energy splitting
of two resolvents in the triangular lattice (three in the rhombic)
with total spin differing by 1. The relevant energy scale is $J=6t^{2}/U$.
To probe this feature further, we calculated its evolution as the
temperature is lowered. As is evident from Fig. \ref{fig:DOS_HF_TEMP},
as the temperature is lowered, the dip-like feature remains below
the chemical potential and more importantly the peak at the chemical
potential sharpens, as would be expected for a metallic state. Consequently,
using the dip-like feature in the density of states as the criterion
for the existence of a pseudogap, we conclude that no pseudogap exists
for the Hubbard model on a triangular lattice for either electron
or hole doping.

However, this analysis is incomplete. Let us approach the pseudogap
problem from a different perspective. In the undoped cuprates, the
quasiparticle dispersion below the Mott gap is characterized by four
maxima at the $(\pm\pi/2,\pm\pi/2)$ points in the Brillouin zone,
as revealed by ARPES measurements on $Ca_{2}CuO_{2}Cl_{2}$\citep{Damascelli_RMP}.
This feature is also present in the half-filled Hubbard model on a
square lattice\citep{Stanescu_Fermi_Arcs}. For a weakly (hole) doped
system, if one adopts the naive picture of a rigid band shift, one
would expect the chemical potential to move to the top of the lower
Hubbard band and intersect it somewhere in the vicinity of the four
minima. The resulting Fermi surface would consist of four small hole
pockets in the vicinity of the $(\pm\pi/2,\pm\pi/2)$ points, while
the rest of the large Fermi surface observed in optimally doped and
overdoped cuprates (or at large doping values in the calculations
for the Hubbard model) would be completely obliterated. That is, the
low-energy excitations are gaped everywhere in the Brillouin zone,
except on the boundary of the small Fermi pockets. This picture seems
to be consistent with ARPES measurements on underdoped cuprates \citep{Shen:2005lp},
as well as the infrared Hall effect \citet{Hall_Cuprates_Rigal,Hall_Cuprates_Shi}
and quantum oscillation measurements \citep{Quantum_Oscillation_Cuprates}.
Nonetheless, strong coupling calculations show that the naive rigid
band picture is, in fact, incorrect and that strong correlations play
a crucial role in  pseudogap physics. 

\begin{figure}
\includegraphics[width=0.5\columnwidth]{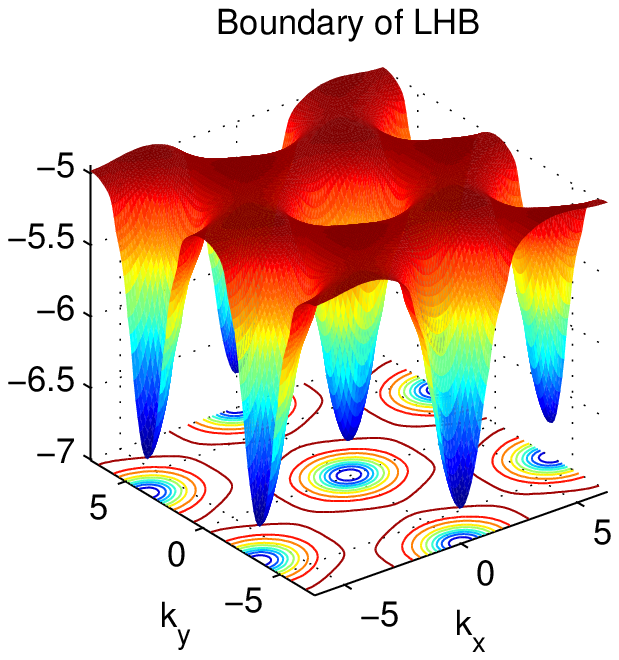}\includegraphics[width=0.5\columnwidth]{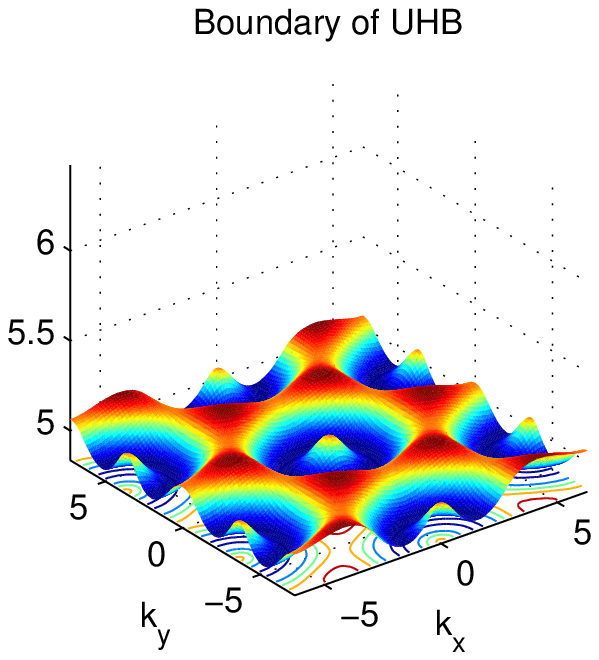}

\caption{The boundary of the lower (left panel) and upper Hubbard band (right
panel) for $U=12t$ and $T=0.1t$. \label{fig:3DPlot_UHB_LHB}}

\end{figure}

One of the essential aspects of strong correlations is spectral weight
transfer. To illustrate its role, let us approach the pseudogap problem
for Hubbard model on a triangular lattice starting from the Mott insulating
phase. Note that, in contrast to the square lattice model, in this
case particle-hole symmetry is always absent and the antiferromagnetic
interactions are frustrated. In Fig. (\ref{fig:3DPlot_UHB_LHB}) we
show the top of the lower Hubbard band (left panel) and the bottom
of the upper Hubbard band (right panel) for a half-filled system with
$U=12t$. The two bands are separated by a Mott gap of about $5t$,
and the chemical potential sits in the middle of the gap. The two
surfaces are defined by the smallest frequencies at which the spectral
function exceeds a certain small threshold $\delta A$. Variations
of $\delta A$ produce only small shifts of the two surfaces, but
their shapes remain essentially the same. First, let us focus on the
upper Hubbard band. As shown in Fig. (\ref{fig:3DPlot_UHB_LHB}),
it is characterized by a set of minima along a large closed curve
around the $\Gamma$ point, not far from the Fermi surface corresponding
to that of the non-interacting half-filled system. A small electron
doping would move the chemical potential near this set of minima of
the upper Hubbard band. In the rigid band picture, one ends up with
two almost circular Fermi surfaces that define a narrow electron ring.
However, our strong-coupling calculation leads to a different picture.
Shown in Fig. (\ref{fig:PseudogapDisc_SF_electron_doping}) is the
spectral function along the $\Gamma\rightarrow K\rightarrow M\rightarrow\Gamma$
path in the Brillouin zone and a small frequency window about the
chemical potential for two small doping values. For comparison, we
also show the bottom of the upper Hubbard band for the insulator (upper
panel) within a similar frequency window. Note that for the insulator
the bottom of the band corresponds to some middle points between $\Gamma$
and $K$ and between $M$ and $\Gamma$. In the vicinity of the $\Gamma$
point, there is no spectral weight within our frequency window {[}see
also Fig. (\ref{fig:3DPlot_UHB_LHB})]. However, once we dope the
system, some spectral weight is transferred from high energies to
low energies, so that a low energy band clearly forms in the vicinity
of $\Gamma$ just below the chemical potential. As a result, the strongly
correlated narrow band that controls the low-energy physics disperses
across the chemical potential generating a large Fermi surface consistent
with the Luttinger theorem. A crucial difference from the square-lattice
case is that the anisotropy along this large Fermi surface (or along
the line defining the minima of the upper Hubbard band for the insulator)
is very weak. By contrast, for the square lattice there is a qualitative
difference between the $(\pi/2,\pi/2)$ and the $(0,\pi)$ regions
of the Brillouin zone as illustrated, for example, by the existence
of minima near the nodal points. This lack of anisotropy leads to
the sudden appearance of a large Fermi surface upon doping and thus
to the absence of a pseudogap.

\begin{figure}
\includegraphics[width=1\columnwidth]{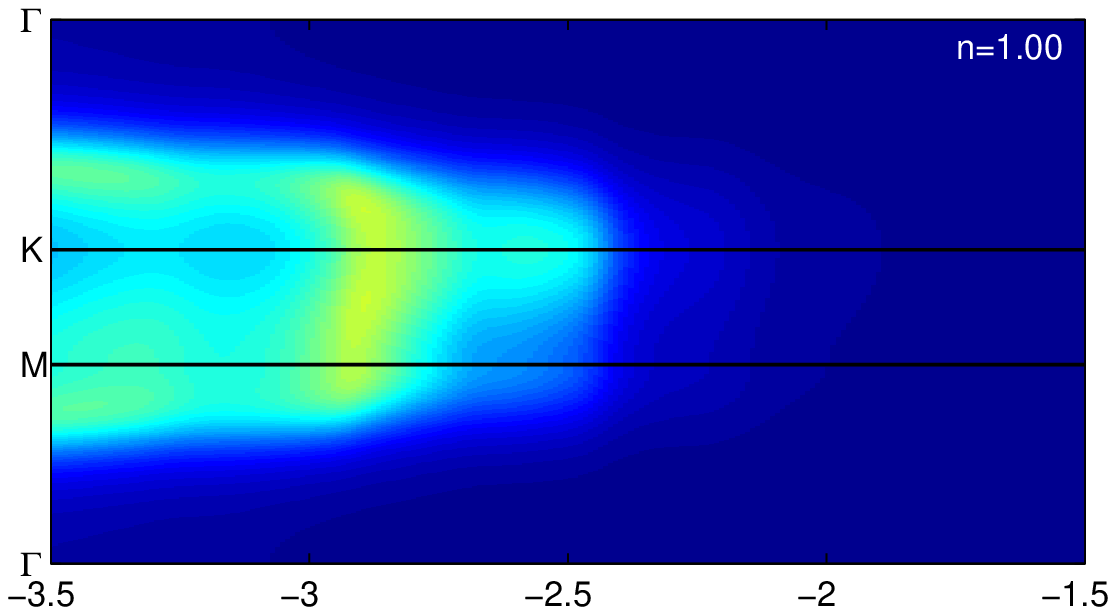}

\includegraphics[width=1\columnwidth]{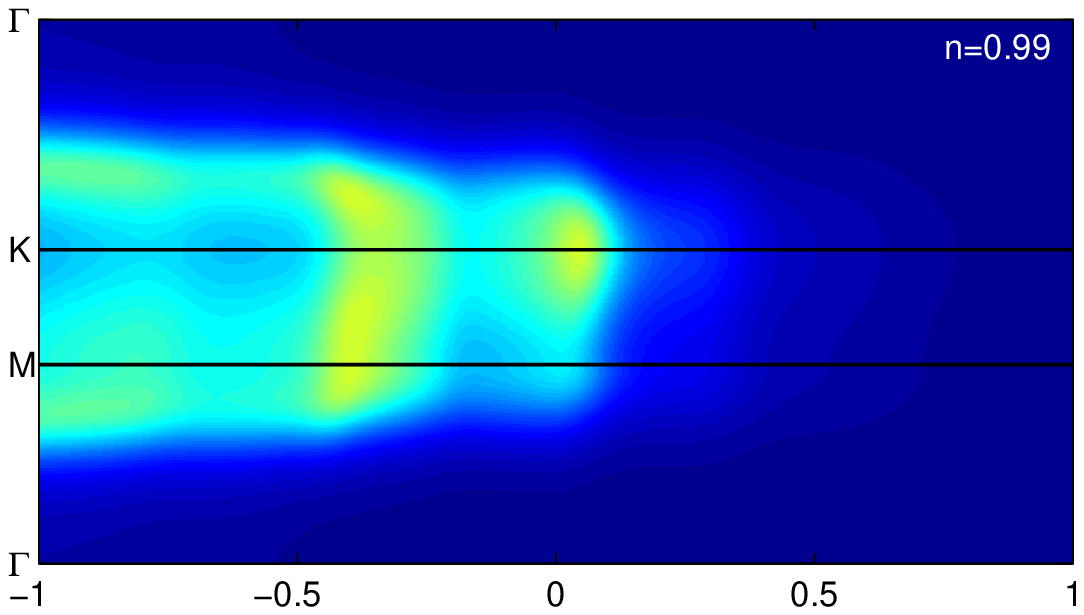}

\includegraphics[width=1\columnwidth]{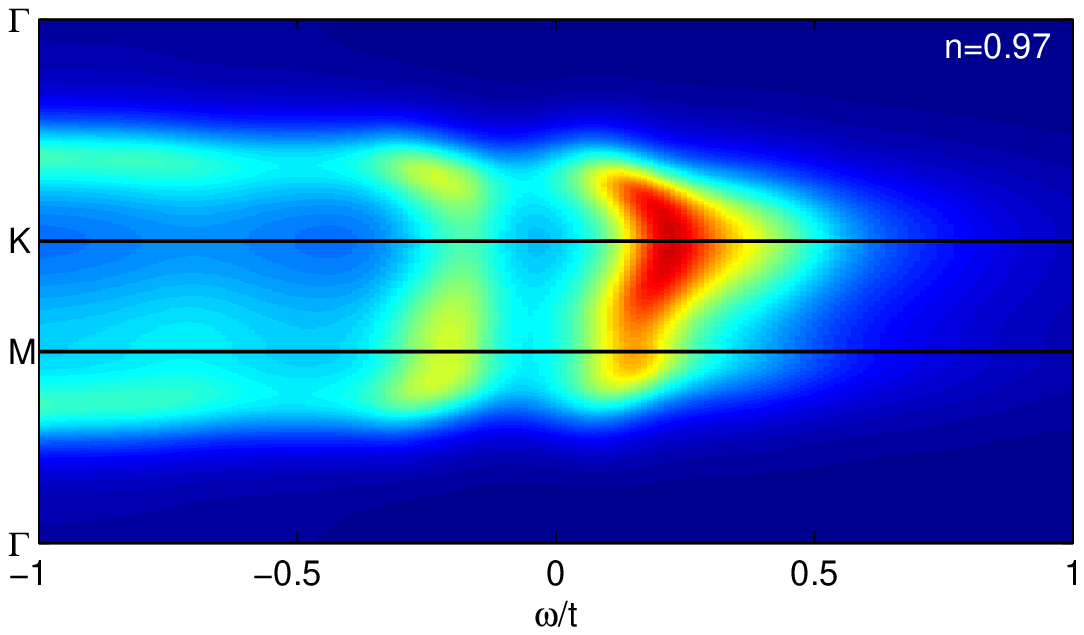}

\caption{Comparison of the spectral function at half filling (top), 1\% (middle)
and 3\%(bottom) hole doping, for the rhombic cluster at $U=12t$,
$T=0.1t$. \label{fig:PseudogapDisc_SF_hole_doping}}

\end{figure}

\begin{figure}
\includegraphics[width=1\columnwidth]{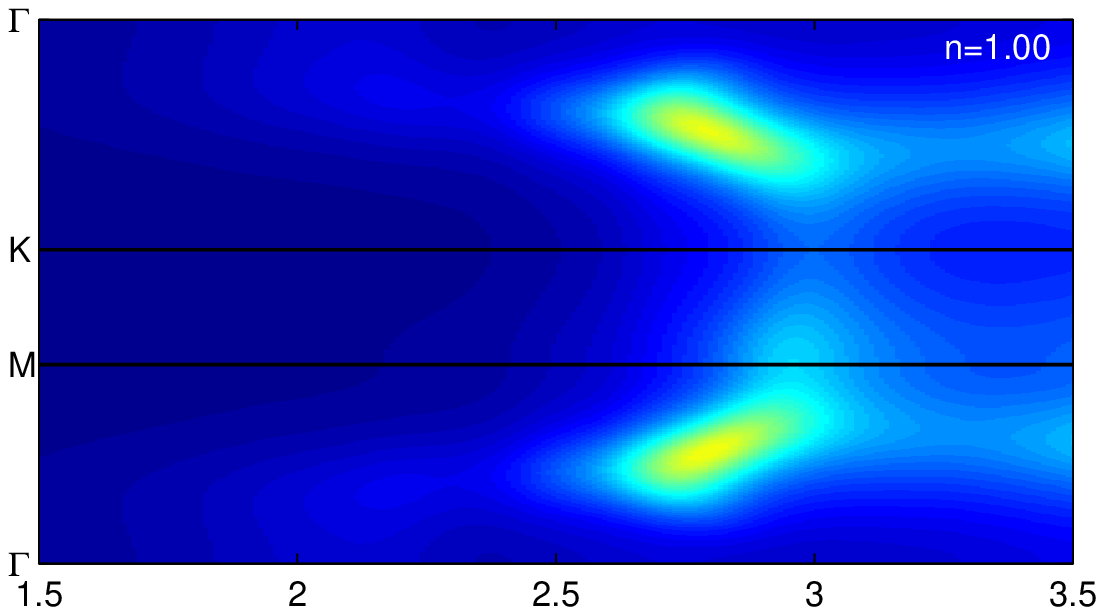}

\includegraphics[width=1\columnwidth]{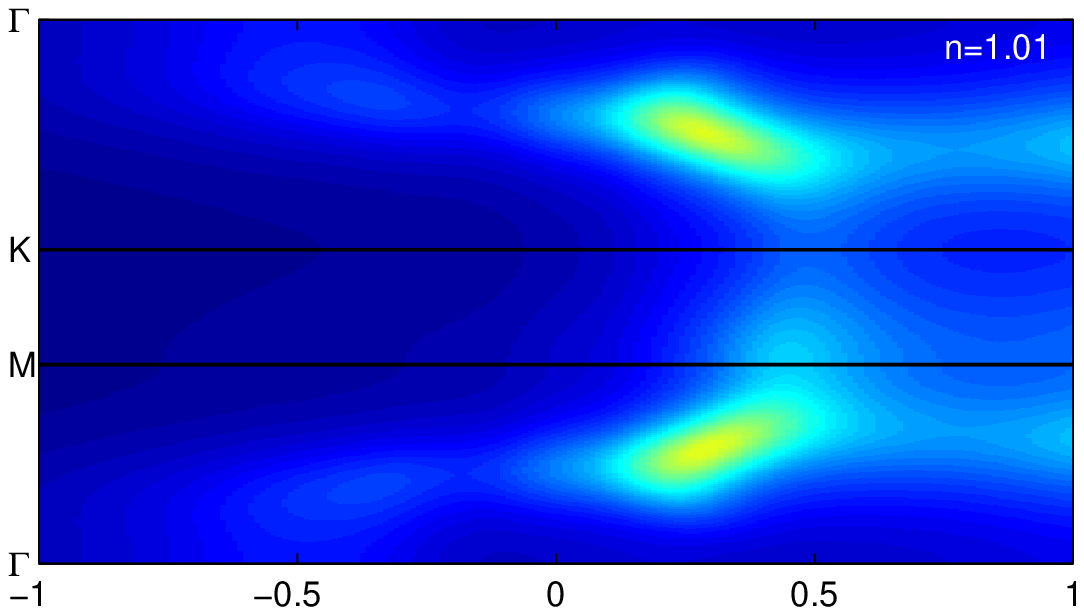}

\includegraphics[width=1\columnwidth]{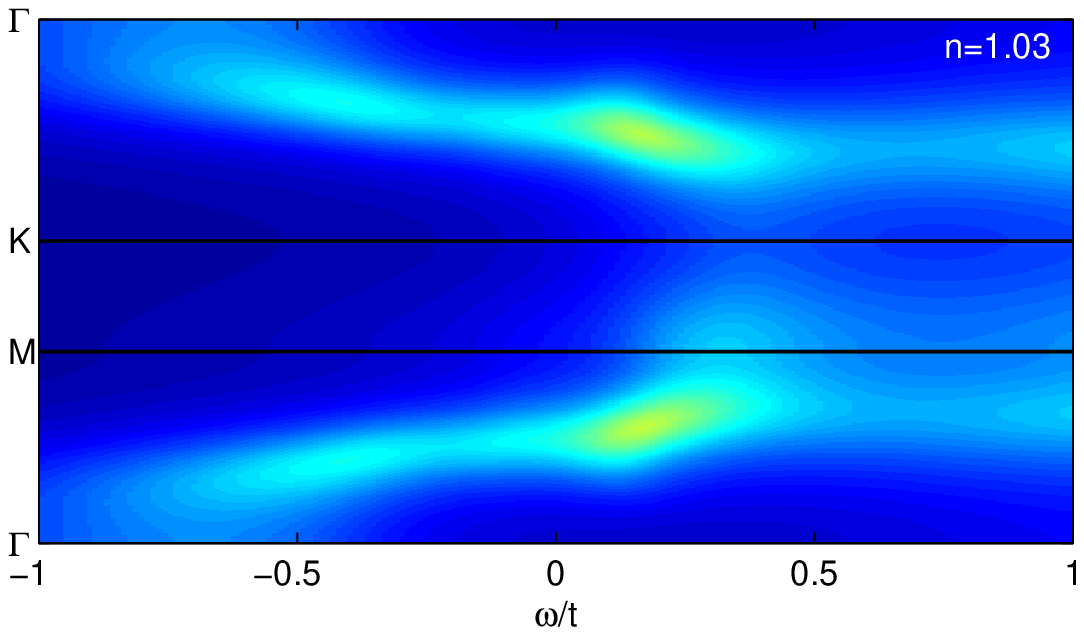}

\caption{Comparison of the spectral function for half filling (top), 1\% (middle)
and 3\%(bottom) electron doping, for the rhombic cluster at $U=12t$,
$T=0.1t$. \label{fig:PseudogapDisc_SF_electron_doping}}

\end{figure}

The situation appears somehow different in the case of the lower Hubbard
band. As shown in Fig. (\ref{fig:3DPlot_UHB_LHB}), the top of the
lower Hubbard band is extremely flat and extends over a significant
portion of momentum space, all around the boundary of the Brillouin
zone. Very weak maxima can be identified near the K points. In this
case, the rigid band picture would suggest that a weakly hole-doped
system is characterized by low-energy excitations that extend over
a large portion of the Brillouin zone and that small Fermi pockets
would possibly form near the K points and extend rapidly with doping.
Again, in the vicinity of $\Gamma$ there is no spectral weight at
low energy. However, in contrast to the upper Hubbard band, this lack
of low-energy excitations at small momenta persists upon doping. Shown
in Fig. \ref{fig:PseudogapDisc_SF_hole_doping} is the low-energy
spectral function along the same $\Gamma\rightarrow K\rightarrow M\rightarrow\Gamma$
path for the insulator (upper panel) and two values of doping. The
relevant spectral weight transfer contributes this time to the re-shaping
of the low-energy narrow band that exists at momenta far from the
$\Gamma$ points. At increased doping values, this band becomes more
dispersive and generates a large Fermi surface consistent with the
non-interacting Fermi surface of a system with the same filling factor.
Nonetheless, at very small doping values, several questions remain.
First, it seems that the chemical potential crosses the narrow band
in an extended area of the Brillouin zone, rather than along a well-defined
Fermi line. This stands in sharp contradiction with Fermi liquid theory.
However, one has to take into account that our results are obtained
at a finite temperature of order $0.1t$, and thus the energy resolution
is severely limited. To establish exactly the position of the Fermi
surface at low dopings would require a much better energy resolution,
and consequently a much lower temperature, would be necessary. The
second question concerns the existence of a pseudogap. One typically
understands the pseudogap as a reduction in the number of low-energy
modes below a certain energy scale. Here we propose a slightly different
view. We define the pseudogap phase as a physical state occurring
close to a Mott insulating state and characterized by the existence
of small Fermi pockets with an area proportional to the doping level
$x=1-n$. By contrast, a normal Fermi liquid is characterized by a
large Fermi surface with an area that is related, via the Luttinger
theorem, to the filling n. Consequently, one should view a system
in the pseudogap phase as a doped Mott insulator. At the same time,
the system represents a re-normalized Fermi liquid characterized by
a reconstructed Fermi surface. From this perspective, the weakly hole-doped
Hubbard model on a triangular lattice is in the pseudogap state. Within
the energy and momentum resolution of the present method, the Fermi
surface appears as a set of small pockets around the K points that
expand rapidly upon doping. A normal Fermi liquid is established at
a doping level of a few percent. We emphasize that a crucial condition
for the realization of this pseudogap phase was the existence of the
small anisotropy in the lowest energy excitations of the Mott insulator.
To study in detail the formation and the evolution of the hole pockets,
calculations using larger clusters (i.e., having a better momentum
resolution) and lower temperatures are necessary. 

\section{Conclusions\label{sec:Conclusions}}

We have studied a strongly correlated electron system on a triangular
lattice using an implementation of the NCA+CDMFT scheme. The key technical
aspects of this implementation are presented in detail. Numerical
results are obtained for two types of clusters containing three and
four sites, respectively. We stress that the cluster size analysis
is a required step in any cluster DMFT-type calculation and argue
that the relative invariance of the result against increasing the
cluster size is the ultimate consistency criterion. The fundamental
issue concerns the short-versus long-range character of the electron
correlations and the nature of the quantity that properly describes
them. We find that the self-energy is not a short-range quantity in
the vicinity of half-filling and therefore cannot be captured using
the cluster components. However, within our momentum and energy resolution,
we find that the cumulant satisfies the locality requirements and
can be used for re-constructing the lattice quantities. In this context,
a very high priority for future cluster DMFT studies should be to
establish the relevant range for the self-energy and the cumulant
in various parameter regimes. Larger cluster calculations are required
to clarify this point. Nonetheless, the task is of pivotal importance
because, if in a certain regime, both the self-energy and the cumulant
are long-ranged quantities, the presently available real- and momentum-space
cluster DMFT schemes are not applicable.

At low doping values, we find that the Hubbard model on a triangular
lattice is strongly correlated with low-energy physics controlled
by a quasi-dispersionless band. As a result of correlations, the band
is very narrow, and its spectral weight can be transferred over large
energy scales. A band with such features cannot be described by any
weakly coupled approach. We also find that a metal-insulator transition
occurs at a critical value of the on-site interaction $U_{c}\approx5.6\pm0.15t$,
which depends very weakly on the size of the cluster. This value is
much lower than the critical interaction determined of $U_{c}\approx10.5t$
in CDMFT calculations using exact diagonalization as the impurity
solver \citet{kyung_CDMFT_Triangle}, but it is closer to $U_{c}\approx6.9t$,
which is the critical interaction obtained by continuous-time Monte
Carlo \citet{lee_monien_dual_fermion}. Finally, we discussed the
pseudogap problem in the context of the Hubbard model on a triangular
lattice. In contrast to the square-lattice case, we find no evidence
for a dip in the density of states positioned at the chemical potential.
However, a momentum-resolved analysis shows that the locus of the
low-energy excitations of the weakly hole-doped system is qualitatively
different from that of a non-interacting system. Therefore, we propose
a new framework for discussing the pseudogap phenomenon, which in
essence involves a momentum-dependent characterization of the low-energy
physics, rather than a momentum-integrated one. We define a pseudogap
state as a state characterized by low-energy excitations occurring
only in a relatively small region in momentum space, qualitatively
different from the location of the low-energy quasiparticles of the
non-interacting system, and having an area that shrinks to zero when
approaching the Mott insulator. Consequently, the system in the pseudogap
state is characterized by a re-constructed Fermi surface consisting
of small pockets. We find that the conditions necessary for the appearance
of these pockets is a strongly momentum-dependent self-energy which
produces quasiparticles with anisotropic properties along the Fermi
surface. Therefore, the pseudogap is intrinsically linked to Mott
physics as emphasized recently \nocite{ftm1,ftm2}, which is the source
of the long-range self-energy. Within the resolution of the present
calculation, we find that the momentum-dependence of the self-energy
is much weaker for the triangular lattice, as compared to the square
lattice, leading to a pseudogap only in the very weak hole-doped regime.
\begin{acknowledgments}
This work was supported in part by NSF DMR-0605769.
\end{acknowledgments}

\end{document}